\documentclass[fleqn,usenatbib]{mnras}

\usepackage{newtxtext,newtxmath}

\usepackage[T1]{fontenc}
\usepackage{ae,aecompl}

\usepackage{graphicx}	
\usepackage{amsmath}	
\usepackage{amssymb}	
\usepackage{xfrac}
\usepackage{subfigure}

\usepackage{comment}

\title[Plasma Structures at 67P]{Plasma Density Structures at Comet 67P/Churyumov-Gerasimenko}

\author[I. A. D. Engelhardt et al.]{
	I.A.D. Engelhardt,$^{1,2}$\thanks{E-mail: ilka.engelhardt@irfu.se}
	A.I. Eriksson,$^{1}$
	G. Stenberg Wieser,$^{3}$
	C. Goetz,$^{4}$
	M. Rubin$^{5}$ \newauthor
	P. Henri,$^{6}$
	H. Nilsson,$^{3}$ 
	E. Odelstad,$^{1,2}$ 
	R. Hajra,$^{6}$ 
	and
	X. Valli\`{e}res$^{6}$
	\\
	$^{1}$ Swedish Institute of Space Physics, Box 537, SE-751 21 Uppsala, Sweden \\ 
	$^{2}$ Department of Physics and Astronomy, Uppsala University, Box 516, SE-75120, Sweden \\ 
	$^{3}$ Swedish Institute of Space Physics, Box 812, SE-981 28 Kiruna, Sweden \\ 
	$^{4}$ Institute for Geophysics and Extraterrestrial Physics, TU Braunschweig, D-38106 Braunschweig, Germany \\ 
	$^{5}$ Space Research and Planetary Sciences, University of Bern, CH-3012 Bern, Switzerland \\
	$^{6}$ LPC2E, CNRS, Universit\'{e} d'Orl\'{e}ans, F-45100 Orl\'{e}ans, France \\
}

\date{Accepted March 20, 2018. Received March 7, 2018; in original form February 7, 2018.}

\pubyear{2017} %

\begin{document}
	\label{firstpage}
	\pagerange{\pageref{firstpage}--\pageref{lastpage}}
	\maketitle
	
	\begin{abstract}	
		We present Rosetta RPC case study from four events at various radial distance, phase angle and local time from autumn 2015, just after perihelion of comet 67P/Churyumov-Gerasimenko. Pulse like (high amplitude, up to minutes in time) signatures are seen with several RPC instruments in the plasma density (LAP, MIP), ion energy and flux (ICA) as well as magnetic field intensity (MAG). 
		Furthermore the cometocentric distance relative to the electron exobase is seen to be a good organizing parameter for the measured plasma variations. The closer Rosetta is to this boundary, the more pulses are measured. This is consistent with the pulses being filaments of plasma originating from the diamagnetic cavity boundary as predicted by simulations.
	\end{abstract}
	
	
	
	
	\section{Introduction}
	
	The Rosetta mission is the first of its kind. The spacecraft closely followed and orbited comet 67P/Churyumov-Gerasimenko (hereafter called comet 67P) for about 2 years (August 2014 to September 2016). The recorded data covers heliocentric distances between 1.25 and about 4~AU (see for example \citet{Taylor2017} for an overview of the mission) as the comet moved along its orbit towards and away from the sun. This was an excellent opportunity to follow the evolution and dynamics of the cometary coma, including its plasma component.
	
	During the whole Rosetta mission Comet 67P was less active than 1P/Halley when it was visited by Giotto at perihelion in 1986 \citep{Almeida2009, Hansen2016}. At 67P collisions of the charged particles with neutral gas molecules therefore are less frequent. Electron cooling on the neutral gas then is less efficient and warm electrons (around 5-10~eV) dominate the electron flux in the inner coma. This can be seen from the spacecraft potential which typically was at least 5~V negative for plasma densities above about 100~$cm^{-3}$ \citep{Odelstad2015,Odelstad2017}. Electrons at energies up to several 100~eV are regularly observed, with tails on the energy distribution up to 10~keV \citep{Broiles2016}. Nevertheless, cooler electrons (below 0.1~eV) were occasionally picked up by the Rosetta Langmuir probe instrument \citep{Eriksson2017}, sometimes dominating the plasma density. Because the energy distribution of recently ionized photoelectrons is flat and wide up to tens of eV \citep{Vigren2013}, the cold electrons must have undergone collisional cooling in the innermost coma. 
	
	The limiting distance outside of which electrons are no longer collisional, and therefore the cooling is inefficient, is known as the electron exobase, also called the electron collisionopause or electron cooling boundary \citep{Mandt2016,Eriksson2017,Henri2017}. This is not a sharp boundary but a region of gradual transition and the exobase distance can be seen as a characteristic scale length. It is defined as the distance to the comet where the neutral gas density scale height is equal to the electron mean free path. This is the same definition that is used for the exobase in a planetary atmosphere where the scale height is that of hydrostatic equilibrium, $H = KT/mg$. However, in the expanding comet atmosphere the neutral gas density decays as $1/r^2$ so the scale height at distance $r$ is $H \sim r$ \citep{Hansen2016}. It can be noted that while the atmosphere is not spherically symmetric, the expansion is still expected to be radial and the neutral gas density depends on the distance as $1/r^2$ \citep{Tenishev2008}.
	
	Inside the exobase the electrons are assumed to lose most of their energy due to collisions with the neutrals. The electron pressure inside is therefore small, while the electron pressure outside this boundary remains high. The exobase distance $L_c$ can be expressed in terms of the neutral density $n_n$ at any given cometocentric distance $r$ as \citep{Eriksson2017} 
	\begin{equation}
		L_c = {n_n \sigma r^2},
		\label{eq:boundary}
	\end{equation}
	where $\sigma = 5\cdot10^{-20}m^2$ is the electron-neutral cross section for 5~eV electrons with water molecules as used by e.g.\ \citet{Mandt2016} and \citet{Henri2017}. To relate observations to how collisional the electrons are at the spacecraft position $r$ it is suitable to give this position in units of $L_c$ as
	\begin{equation}
		R^* = \frac{r}{L_c} = \frac{1}{n_n \sigma r}.
		\label{eq:disboundary}
	\end{equation}
	We can note that Rosetta orbits most of the time outside this boundary, $R^*>1$ \citep[Fig.\ 5]{Mandt2016}. This does not mean that the electrons are completely collisionless, but a higher value of $R^*$ indicates less local collisionalilty.
		
	Not only is the comet weakly active and the plasma environment not fully developed, but the plasma environment also turns out to be very unstable. The most prominent example is the detection of the 'singing comet waves', with frequencies of 10 -- 100~mHz and very large amplitude, d$B/B \sim$1 \citep{Richter2015,Koenders2016a} in the low activity stages of the mission, but strong variations in all plasma parameters are found during all the mission. 
	
	The plasma density turns out to be very variable in all regions investigated by Rosetta \citep{Edberg2015,Odelstad2015,StenbergWieser2017}. It is smoother inside the diamagnetic cavity \citep{Goetz2016,Goetz2016b,Henri2017}, but large density fluctuations have been observed there as well \citep{Hajra2017}. Outside the diamagnetic cavity, pulses of higher density are regularly observed. They vary in duration but are typically on the order of a few to a few tens of seconds in the examples shown by \citet{Eriksson2017}. 
	
	The boundary of the diamagnetic cavity seems to be unstable. Hybrid simulations by \citet{Koenders2015} predicted that filaments of plasma are cut off from the diamagnetic cavity and move into the magnetized region outside, with the instability being most prominent in the plane perpendicular to the external magnetic field. These then are sharp density increases, much like the pulses we see, or filaments that move tailward. Rosetta observations also show that many cavity observations are short, around a minute or even less, and \citet{Henri2017} suggests that these observations can be interpreted as finger-like structures extending out from a central cavity into the surrounding magnetized plasma. It can be noted that the filaments found by \citet{Koenders2015} in the hybrid simulations do not have zero magnetic field, and so cannot be directly identified with the cavity fingers suggested by \citet{Henri2017}. \citet{Henri2017} also show that the occurrence statistics of these brief cavity observations are well organized by $R^*$, pointing at the importance of the electron collisionality for the cavity physics.
	
	In this paper, we investigate the pulses of high plasma density observed outside the diamagnetic cavity by \citet{Eriksson2017} while \citet{Hajra2017} studied plasma density pulses inside the diamagnetic cavity. These pulses, where the density is higher than the background density have a relatively short duration, typically a few seconds to a few tens of seconds, as seen from the spacecraft. However, longer durations (up to about 10~minutes) have been observed. As shown by \citet{Eriksson2017} they are very common, and it is therefore important to understand them and their role in the comet plasma environment. Four events are discussed in detail (Section~\ref{sec:obs}), based on data from several Rosetta instruments (Section~\ref{sec:data}). To investigate their relation to the electron cooling, we also do a statistical investigation of their occurrence (Section~\ref{sec:boundary}).

	\section{Methods and Instruments}
	
	\subsection{RPC-LAP}
	
	The main instrument used in this study is the RPC (Rosetta Plasma Consortium \citep{Carr2007}) LAP (LAngmuir Probe) instrument \citep{Eriksson2006}. It uses two separate spherical Langmuir probes (LAP 1 and LAP 2) that are identical and can be operated in different modes. They are mounted on booms (2.2~m and 1.6~m, respectively) that separates the probes by 5~m.
	
	LAP can be put into different modes for accessing different parameters and for adopting the measurements to the plasma conditions and available telemetry. The main modes are (1) bias voltage sweep, (2) constant bias potential and (3) constant bias current, including floating probe, mostly used for setting both probes together in an E-field mode (see \citet{Eriksson2006} and \citet{Eriksson2017} for details). The mode used in this study is when the probe is set to a constant bias potential, attracting either electrons or ions (2).
	As the currents carried by plasma particles to a Langmuir probe are proportional to the plasma density, this mode is useful for high time resolution measurements of plasma density dynamics. We will also use LAP modes with at least one probe measuring voltage when floating, i.e.\ not exchanging any current with the surrounding plasma, providing a measure of the spacecraft potential \citep{Odelstad2017}.
	
	The operational modes are defined and operated by so called macros. These are short scripts that run the instrument in the appropriate mode. A macro is typically run for a time span from a few hours to about a day until a new macro is commanded. The interval in which a particular macro runs is called a macro block. The macro blocks are the basic divisions in time we use in this study.

	To get as complete information on the pulses as possible, we will present data from several events (see Section~\ref{sec:events}) with LAP in different operational settings, measuring different quantities.
	The macros mainly used here are internally identified as 624 and 914. We will refer to them by the more descriptive designations EI (for electron-ion) and II (for ion-ion), respectively. In macro EI, LAP 1 is set to a bias potential of +30~V and LAP 2 to -30~V, and so the two probes mainly sample electrons and ions, respectively. This is the mode used in the example shown by \citet{Eriksson2017}, where the simultaneous increase in both these currents showed that the pulses observed by LAP are due to real increases in plasma density. The probe currents are continuously recorded at 57.8~Hz except for short data gaps (less than a second) every 32~s and breaks for probe bias sweeps for more complete plasma characteristics at 160~s intervals.
	Macro II is similar to EI but has both probes set to -30~V, which means that the current they measure is due to plasma ions attracted to the probe, with some addition from the almost constant or at least slowly varying photoelectron emission \citep{Johansson2017} from the probe. Since the two probes operate in the same way, the two signals can be directly compared.
	For one event macro 802 is used, hereafter called VV (for voltage-voltage). Here both LAP probes are used in voltage mode with no bias current or voltage applied, and the potential of the freely floating probes is measured at 57.8~Hz. As shown by \citet{Odelstad2017}, this gives a measure of the ()negative) spacecraft potential, $V_S$, which in turn is sensitive to the plasma density. 
	
	Neither of the LAP modes above give information on the absolute value of the plasma density unless complemented by some assumption or measurement of the electron temperature. Such can be derived from LAP in sweep mode \citep{Eriksson2017}. Another way is to use the proportionality of the probe current to plasma density together with the independent plasma density measurement provided by the RPC-MIP instrument for calibration. This will be done in section \ref{sec:calibration}.

	\subsection{RPC-MIP}
	
	The Mutual Impedance Probe (MIP) \citep{Trotignon2007} is also part of the RPC. It transmits a signal at different frequency steps and observes the response at the same frequency. The plasma density is then retrieved by on-ground identification of characteristics, such as the resonance peak, of the mutual impedance spectrum \citet{Gilet2017}.
	MIP is considered to provide a reliable measurement for the total plasma density when a resonance peak can be clearly identified in the mutual impedance spectra. This means in practice that the plasma density has to be above or around 100~cm$^{-3}$ (limit somewhat depending on electron temperature). 
	MIP data (usually available at a time resolution of between 4 and 32~s) can be used to calibrate LAP current sampled at higher rate (57.8~Hz for the data presented here) to a plasma density value using of a linear fit. This also extends the density range to values too low for MIP to measure it. LAP observations can give electron density even if MIP can not. This is further discussed in Section~\ref{sec:calibration}.
	
	\subsection{RPC-ICA}
	
	ICA \citep{Nilsson2007} is the Ion Composition Analyzer of the RPC. It measures the mass-separated energy distribution function of positive ions from a few eV up to 40~keV within $45^\circ$ of the detector plane, so the solid angle coverage is about $2 \pi$~sr. ICA is mounted on the spacecraft so that both the Sun and the comet nucleus are in the field of view of the instrument during nominal pointing when all imaging instruments on Rosetta look toward the nucleus \citep{Nilsson2015}. It usually can be assumed to cover the most important flow directions the ions are expected to come from, i.e. from the nucleus or streaming with an anti-sunward component. The angular distribution of cometary ion flow throughout the mission and detailed examples was shown by \citet{Nilsson2017}. If there is a population of low energy ions and if the spacecraft potential, $V_s$, is negative enough, the ICA ion observations give the spacecraft potential. 
	For a negative $V_s$, all ions will have been accelerated to an energy at least $e V_s$ when they reach ICA, so the lowest energies detected give $V_s$ \citep{Odelstad2017}. In the later part of the mission, ICA often used a mode with limited energy range but high time resolution \citep{StenbergWieser2017}. As this has sufficient time resolution for the pulses we are interested in, this is the mode used in this study.

	\subsection{ROSINA-COPS}
	
	From ROSINA (Rosetta Orbiter Spectrometer for Ion and Neutral Analysis, \citet{Balsiger2007}) we use data from the nude gauge of the COPS (COmet Pressure Sensor). It measures the total neutral gas density. 
	The primary measured quantity is the ion current to a small sensor. However the ions are created by ionizing the neutral gas in a volume designed to only let the gas in and keep the plasma out by electrostatic means. The resulting current is thus proportional to the number density of the neutral gas. Nevertheless, sufficiently energetic plasma particles may enter COPS at times \citet{TzouThesis}, and so we will not consider COPS signatures of fast variation as real neutral gas density variations. Our main use of COPS instead is to provide the large scale neutral gas background, which is important not only as source of the comet ionosphere but also for cooling of the electron gas.
	
	\subsection{RPC-MAG}
	
	MAG, the RPC fluxgate magnetometer \citep{Glassmeier2007,Richter2015}, measures the magnetic field vector. The MAG data used here are sampled at 20~Hz. No noise reduction has been applied to the data, only calibration by removing the mean of the magnetic field in each component inside the diamagnetic cavity, when available. Our chief interest in MAG data is for comparing the magnetic signal to the local density fluctuations, but we also use MAG to see if the phenomena we study are organized by the magnetic field. Due to sensor temperature sensitivity and noise from spacecraft systems, some quasi-constant offsets may sometimes remain in MAG data. This is not a problem for identifying transient structures, but some caution is needed when calculating angles of the magnetic field when the magnetic field is weak.
	
	\subsection{Event Selection}
	
	The main focus of this study is to show details of the strong plasma density fluctuations seen as high density pulses by Rosetta. These were most common in the months around perihelion \citep{Eriksson2017}, so we concentrate on events during the highest cometary activity phase of the Rosetta mission. The events are chosen from Autumn 2015, within a few months after perihelion (Aug 13, 2015). Data from the months leading up to perihelion could also have been used. However, new LAP operational modes tailored to the conditions found around perihelion were uploaded in early August and MIP operations were also optimized in the same time frame, so the post-perihelion phase offers better opportunities. 
	
	We concentrated on periods when all the used RPC instruments operated in their high telemetry rate, known as burst mode, to get high time resolution data. In order to avoid effects of changing probe illumination stable spacecraft pointing was required for all events, with only minuscule changes (few degrees) in attitude. With these constraints at least one example interval was identified for each of the LAP modes EI, II and VV, as each of these can illuminate different aspects of the pulses. 
	
	Rosetta spent most of its time in the terminator plane \citep{Taylor2017}, so selecting events randomly may give the impression that the pulses only exist there. To show that this is not the case, we made sure some events included were from different solar zenith angles (SZA, also known as phase angle). 
	
	The nucleus rotation period is close to 12~hours and the outgassing varies over the comet surface and with illumination. Intervals of 12~hours are therefore desirable, but some shorter intervals had to be accepted. Finally, periods when ICA was operated in its low-energy high-time-resolution mode were preferred, though not all events could be chosen to include such data. Weighting all the requirements resulted in the set of four events to illustrate various features of the pulses listed in Sections~\ref{sec:oct24} -- \ref{sec:nov15}. 
	
	Table~\ref{tab:cases} shows an overview of the events investigated. The neutral density $n_n$ from COPS and the distance $r$ of Rosetta to the center of the nucleus change during the intervals but average values are given. From these values we also calculate an approximate production rate (molecules per second) relevant for the event by assuming spherical symmetry and a gas outflow speed $u = 1$~km/s as $Q = 4 \pi r^2 n_n u$.
	
	\begin{table*}
		\centering
		\caption{An overview of the cases presented in this paper. Given are the date, time, LAP macro, as well as approximate values for the COPS neutral density $n_n$, radial distance $r$, gas production rate $Q$ and solar zenith angle SZA.}
		\label{tab:cases}
		\begin{tabular}{rccccccc}
			\hline
			Date & Blocktime (start)& Macro & $n_n$  & $r$& 	Q	& SZA & Remarks \\
			 	& [h]			&		& [$10^7$cm$^{-3}$] & [km] &	[$10^{27}$s$^{-1}$]	 & [$^o$]& \\ 
			\hline
			Oct 24, 2015 & 06-16 (14) & EI (624)  & 1 &  400 & 20 & 60 &Both probes sunlit \\
			Nov 15, 2015 & 00-12 (08) & II (914)  & 2 &  140 & 5 & 60 & - \\
			Nov 15, 2015 & 12-00 (12) & VV (802)  & 2.5 &  160 & 8 & 60 & E-field mode \\
			Nov 20, 2015 & 20-00 (22) & EI (624)  & 3 &  150 & 8 & 90 & - \\
			\hline
		\end{tabular}
	\end{table*}
	
\begin{figure*}
		\centering
		\subfigure[a]{	
			\includegraphics[width=\columnwidth]{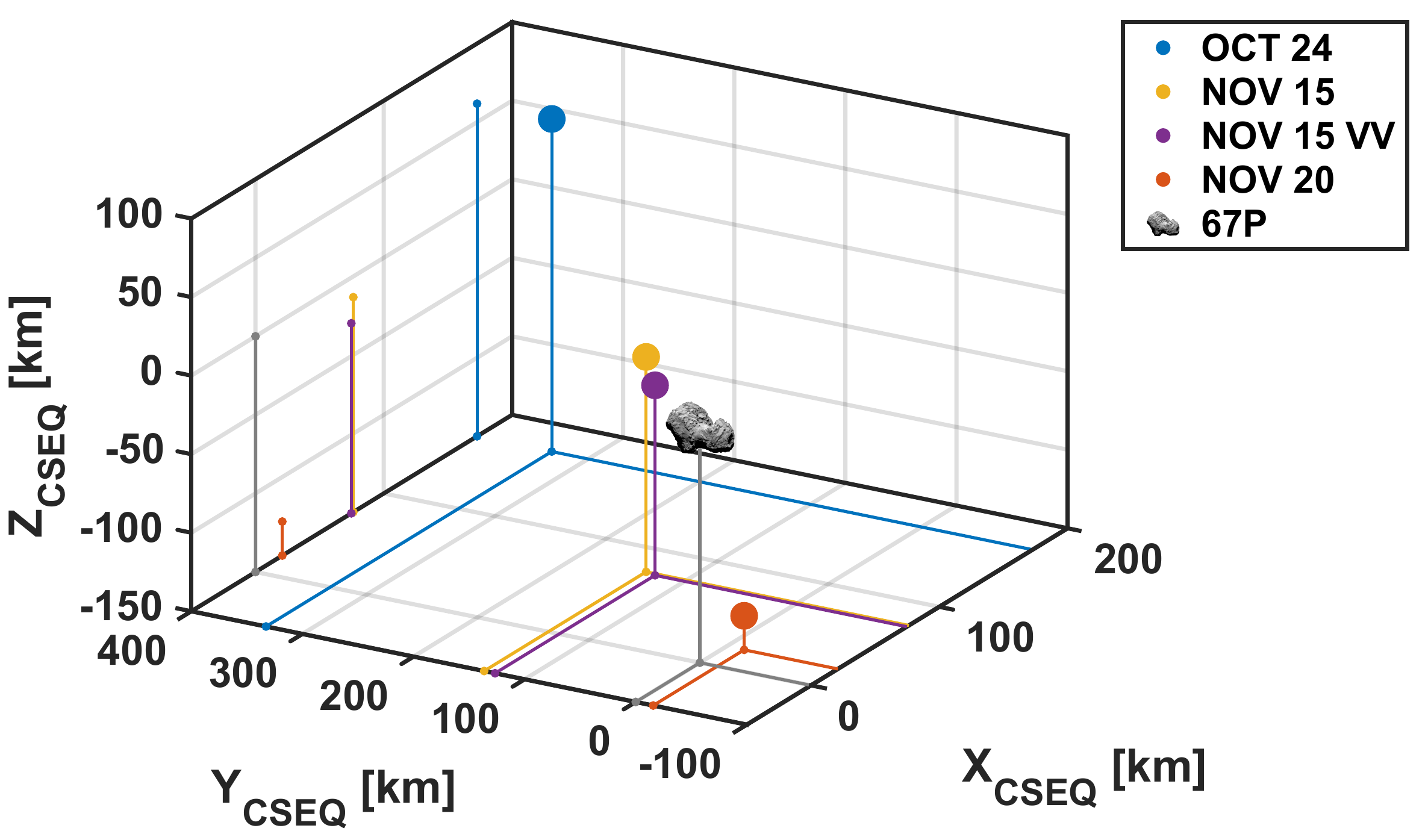}
			\label{fig:Orbit}
		}
		\subfigure[b]{	
			\includegraphics[width=0.6\columnwidth]{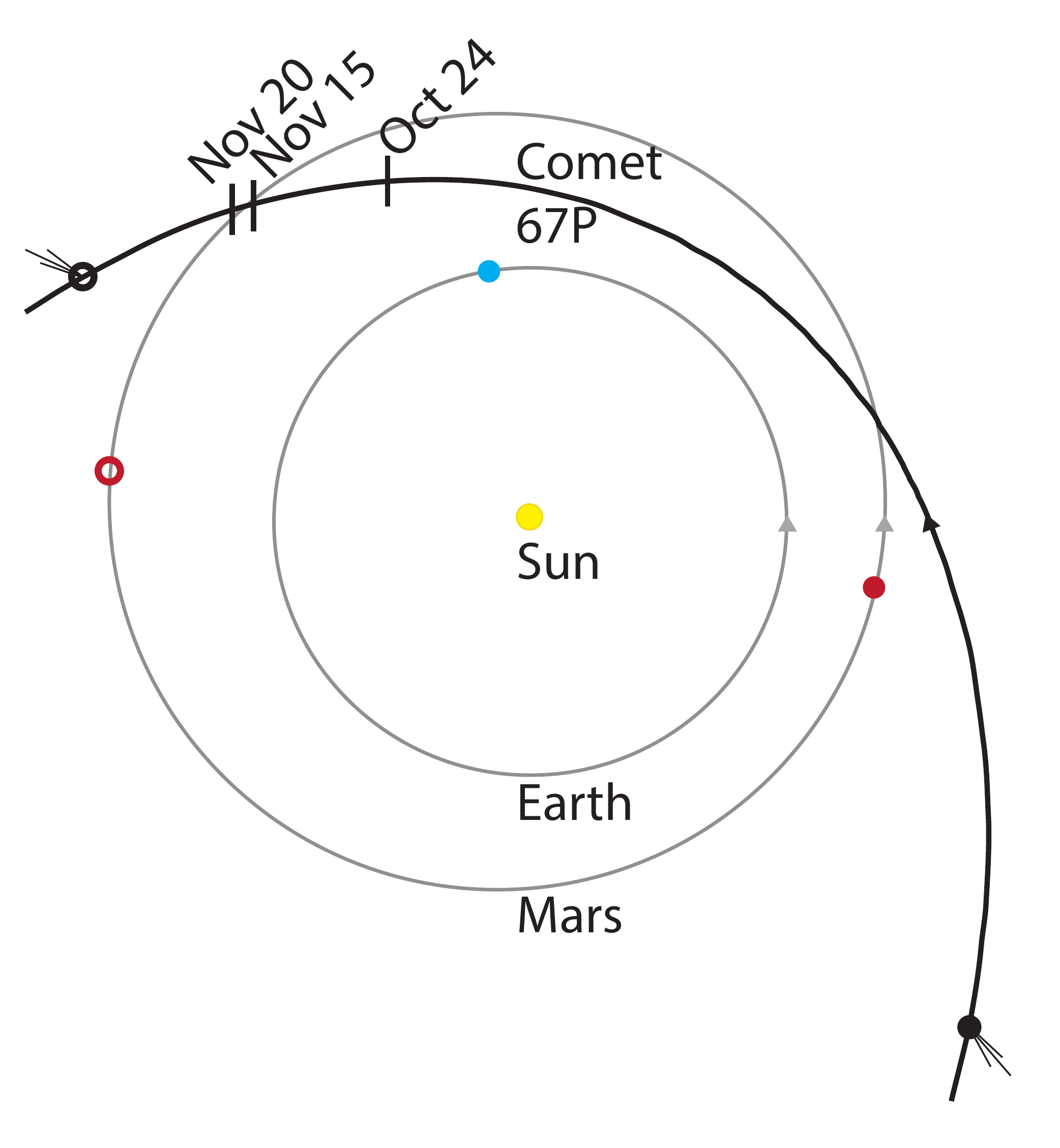}
			\label{fig:orb2015}
		}
		\caption{(a): Position [km] of Rosetta relative to 67P (gray circle) for the events in Table~\ref{tab:cases}, averaged over the macro blocks, given in CSEQ coordinates. This is the cometocentric solar equatorial coordinate system. Here the x-axis points from the comet to the Sun, the z-axis is the component of the Sun's rotation vector, perpendicular to the x-axis, and the y-axis completes the right-handed reference frame (e.g. described by\citep{Edberg2016}). Fig. (b): Rosetta, Earth and Mars positions during 2015 seen from 90$^o$ above the ecliptic plane, with the position of the selected events marked. Filled circle is the starting position and empty circle is the final position. 
		}
	\end{figure*}

	Figure \ref{fig:Orbit} shows the 3D position of Rosetta at the times of the events we have investigated. The coordinate system used is CSEQ. The comet nucleus is marked by the gray circle at the origin. It can be observed that all observations are taken on the day side, i.e.\ at positions with positive $X$ coordinate. This reflects the operational constraint that the spacecraft should not enter the night side of the nucleus. Two of the events are from the terminator plane ($X_{CSEQ} \approx 0$) and the remaining three from the day side at significant phase angle (solar zenith angle).
	
	\citet{Eriksson2017} showed an overview of the pulse occurrence in LAP data over the full year 2015, spanning 1.25 to 2.5~AU. This period can be seen in figure \ref{fig:orb2015} where full circles mark the position of the various objects at the start of 2015 and the empty circle the end. The picture also shows the approximate position of 67P during the chosen events.

	\subsection{LAP-MIP Cross Calibration}
	\label{sec:calibration}
	
	In difference to MIP, the current (or voltage) detected by LAP primarily depends on the particle fluxes reaching the probe and (for the spacecraft potential) the spacecraft. For a constant shape of the energy distribution, the fluxes and thus the LAP current should only depend on the density \citep{Eriksson2017}, but otherwise variations in e.g.\ temperature also has an impact on LAP. 
	As long as the spacecraft potential is stable, the LAP current will be proportional to the plasma density. 
	The random electron current flowing to a spherical probe at the same potential as the surrounding plasma is proportional to the electron density $n_e$ as given by
	\begin{equation}
		I_{e,o} = e A_{LP} n_e \sqrt{eT_e/2 \pi m_e },
		\label{eq:Ipropn}
	\end{equation}
	where $e$, $T_e$ and $m_e$ are electron charge, temperature and mass, respectively, and $A_{LP}$ is the surface area of the probe.
	A real probe will be at some potential to the plasma, and standard theoretical expressions give a correction factor for this 
	\begin{equation}
		I_e = I_{e,o} \left( 1+ \frac{V_{b}+V_s}{T_e} \right)
		\label{eq:Iefull}
	\end{equation}
	where $V_{b}$ and $V_s$ are the bias and spacecraft potential, respectively \citep{Eriksson2017}. The ion current to a negative probe obeys a similar expression.
	
	To use these expressions for deriving the density from the LAP current we have to assume values for parameters like the spacecraft potential, electron temperature and ion velocity which are not known at as high time resolution and always with some uncertainty. 
	Even though the spacecraft potential $V_s $ is also dependent on the plasma density, we can at least for a limited density range assume that the electron current is proportional to the density as in Equation \ref{eq:Ipropn}. We then apply a fit 
	\begin{equation}
		n_{MIP} = A (I_{LAP} + B).
		\label{eq:nAB}
	\end{equation}
	Since in a plasma the $n_e \sim n_i$ we can use the same formula to find the appropriate fit for ion and electron current.
	To derive the density from the potential measurement, VV mode, we use that the spacecraft potential is related to plasma density as \citep{Odelstad2017,Heritier2017}
	\begin{equation}
		n \propto \exp(-\alpha V_{s}/T_e)
	\end{equation}
	so we can do a linear fit of $\log n$ to the measured probe potential $U_{LAP}$,
	\begin{equation}
		\log n_{MIP} = \log D + U_{LAP}/C.
		\label{eq:nCD}
	\end{equation}
	Table \ref{tab:LapFits} shows the used fits for calibrating LAP current and voltage to plasma density to MIP. 
	As expected, the fit coefficients are different for probes in positive and negative bias potential (E and I modes). We also see a significant difference between events, particularly in E mode. This is due to different plasma conditions which in turn cause different spacecraft potential. This can be seen in eqn.~\ref{eq:Iefull} to change the linear relation of the density and current.

	\begin{table*}
		\centering
		\caption{An overview of the fit parameters resulting from calibrating LAP currents to MIP plasma density, according to equations \ref{eq:nAB} and \ref{eq:nCD}.
		}
		\label{tab:LapFits}
		\begin{tabular}{cccccc} 
			\hline
			Date & Macro & \multicolumn{2}{c}{Fit LAP 1} & \multicolumn{2}{c}{Fit LAP 2} \\
			&		 &	$A$ [cm$^{-3}/$nA]		& $B$ [nA]	 	& $A$	[cm$^{-3}/$nA]		& $B$ [nA]	\\
			\hline
			Oct 24, 2015 & EI (624)  & 0.97 	&  0.6		& 60 		& - 18.3 \\
			Nov 15, 2015 & II (914)  & 20 	& - 10.0 	& 20 		& -10.0 \\
			Nov 20, 2015 & EI (624)  & 0.007 	& 649.3		& 18.5 	& 6.4 \\
			\hline
			&   & \multicolumn{2}{c}{Fit LAP 1 and LAP 2} &  &  \\
			&	 &	$C$ [eV]	& $D$ [cm$^{-3}$]	 & 	&  \\
			\hline
			Nov 15, 2015 & VV (802)  & 3.3 		&  200		&  			& \\
			\hline
		\end{tabular}
	\end{table*}

	\subsection{Data Presentation}
	\label{sec:data}
	
	To provide an overview of the coma environment for the selected events (see Section~\ref{sec:obs}, Fiugres~\ref{fig:Oct24_full} to \ref{fig:Nov152_full}), we present data from LAP, MIP, ICA (where available), COPS and MAG.
	The panels are further explained in the following description.\\
	
	{\bf Panel A:}
	Shows the plasma density by both LAP probes (blue and red) and MIP (orange). The calibration is explained in Section \ref{sec:calibration}.
	
	{\bf Panel B:}
	The ICA high resolution data is plotted, where available. The vertical axis shows the energy in eV and the color scale gives the flux of ions for each energy channel. 

	{\bf Panel C:}
	On the left vertical axis of the plot (blue curve) we show the neutral density as derived from ROSINA COPS. The right vertical axis (red curve) shows the spacecraft position relative to the electron exobase distance, $R^*$ (equations \ref{eq:boundary} and \ref{eq:disboundary}).
	
	{\bf Panel D:}
	This panel shows the magnetic field measured by MAG in CSEQ coordinates. $B_{tot}$ is shown in black, $B_x$ in red, $B_y$ in green, and $B_z$ in blue.
	
	{\bf Panel E:}
	Shows the angle (blue) between the Comet-Rosetta vector and the local magnetic field, for $B_{tot}>10~nT$, as well as the solar zenith angle (green). 
	The red horizontal bar marks angles between 85 and 95 degrees.

	\section{Observations}
	\label{sec:obs}
	\label{sec:events}

	\subsection{October 24, 2015}
	\label{sec:oct24}
	
	\begin{figure*}
		\centering
		\subfigure[10 hours]{
			\includegraphics[width=\columnwidth]{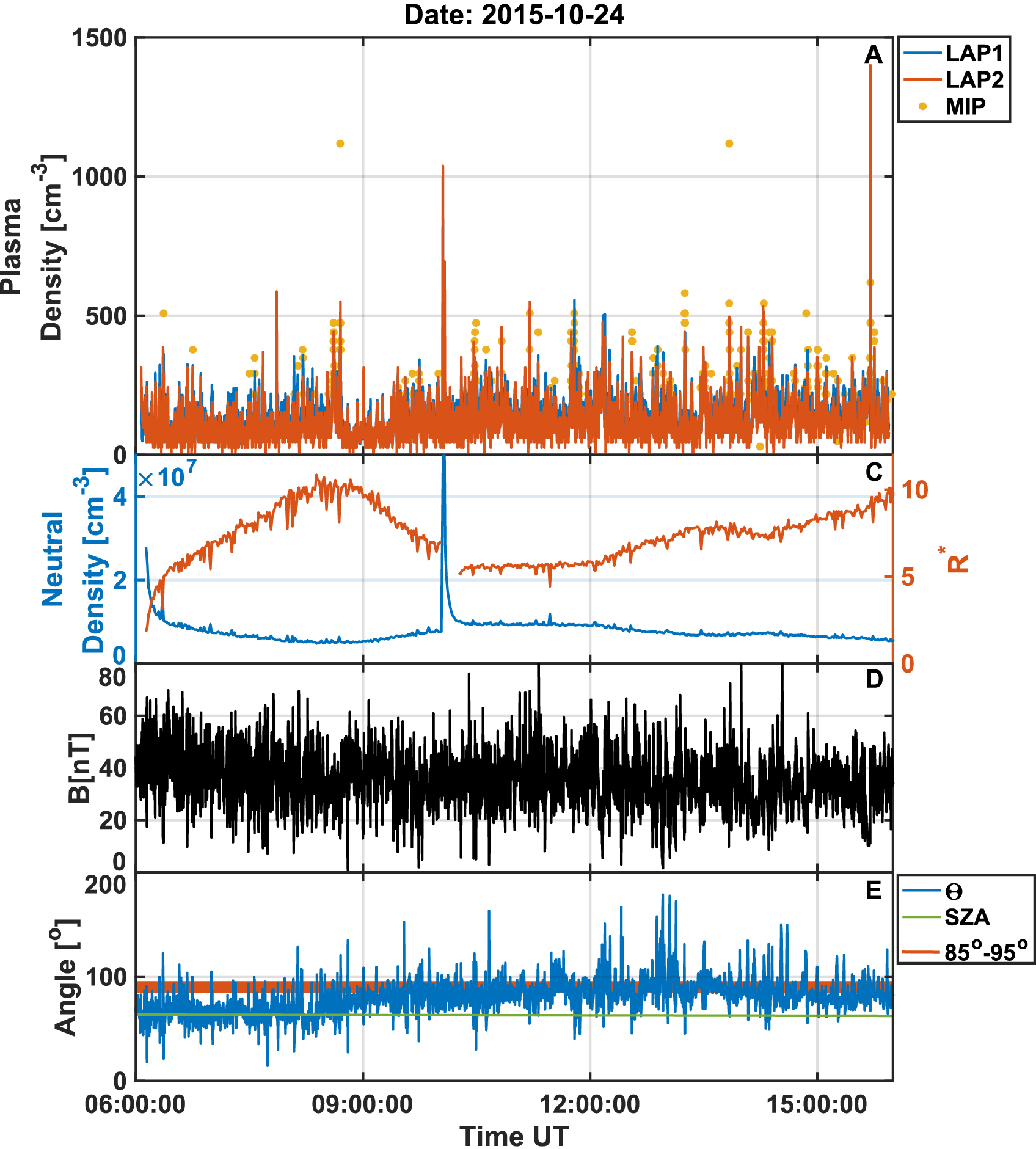}
			\label{fig:Oct24}
		}
		\subfigure[1 hour]{
			\includegraphics[width=\columnwidth]{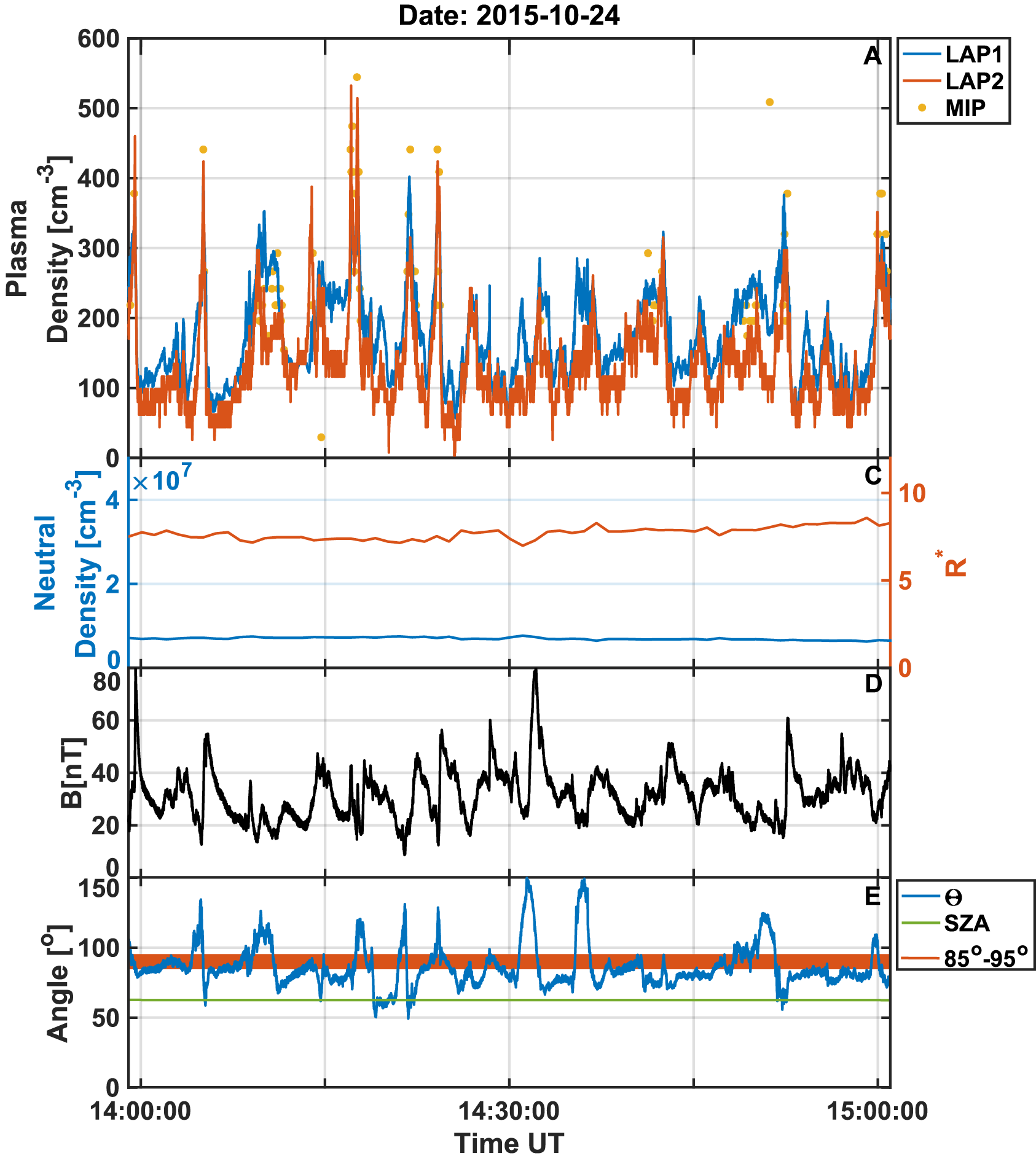}
			\label{fig:Oct24H}
		}
		\caption{Fig.~\ref{fig:Oct24}: Data for October 24. LAP is in EI-mode for this day. Fig.~\ref{fig:Oct24H}: One hour from the October 24 data set (14:00-15:00). The panels in both figures are as described in Section \ref{sec:data}. The data spike in LAP and COPS (panels A and C) around 10:00 is due to a spacecraft thruster firing and should be neglected. }
		\label{fig:Oct24_full}
	\end{figure*}
	
	The first event we present here is October 24, 2015. As seen in Figure~\ref{fig:Orbit}, Rosetta was in the positive $Z$ hemisphere of the comet, on the evening side of the nucleus at a solar zenith angle around 60$^\circ$ and distance of 400~km. Evening (and morning) is here defined in terms of local time at the point on the rotating nucleus directly below Rosetta. Figure\ref{fig:Oct24} shows an overview of the full macro block with LAP in the EI mode. LAP 1 samples electrons and LAP 2 samples ions. Both currents have been fitted to the MIP plasma density as described in Section~\ref{sec:calibration}. The spike seen in LAP (panel~A) and COPS (panel~C) data at around 10:00 is due to a spacecraft thruster firing and should be neglected. We will not go further into effects of these maneuvers to various instruments. We only note that these effects are common and usually decay within a few tens of minutes \citep{TzouThesis}.
	The plasma density in LAP 1 and 2 seem to follow each other well which is more apparent in the zoom in figure \ref{fig:Oct24H}. This figure shows a one hour interval of figure \ref{fig:Oct24}, and here a lot of density spikes show up clearly in both probes. The plasma density, $n$, is a few hundred cm$^{-3}$ inside the pulses while it is less than 100~$cm^{-3}$ outside. MIP is not able to resolve the lower density outside the pulses but together MIP and LAP cover the density range of interest. 
	
	Some simultaneous variations may also be present in the neutral gas (COPS, Panel~B), but they are much smaller (< 10\%) than the plasma signatures (factor of ~5 in LAP current). As it is known \citep{TzouThesis} that COPS may sometimes react to plasma variations we do not interpret these small variations as due to real fluctuations in the neutral gas. The COPS data indicate that the spacecraft is at about 8 times the cometocentric distance of the electron exobase (red line in Panel~C). As diamagnetic cavity observations were mostly found closer to the electron collsionopause \citep{Henri2017}, this also indicates Rosetta is quite far outside the diamagnetic cavity. In fact no diamagnetic cavity boundary crossing was identified by \citet{Goetz2016b} during the whole month of October 2015, since during this month a dayside excursion happened and Rosetta was far away from the nucleus.
	
	The magnetic field (panel~D) generally varies in the same way as the LAP and MIP data, with density and magnetic field strength increasing together, further verifying the plasma nature of the structures. However, the pulses are more asymmetric in the magnetic field than in the density, with a sharp leading edge and a slow decay. This assymmetry is also seen in the ion energy and spacecraft potential \cite{StenbergWieser2017}. As shown by \citep{Hajra2017} in observations of similar structures just outside the diamagnetic cavity, this asymmetry can often be found also in the density but it is not obvious here.

	\subsection{November 20, 2015}
	\label{sec:nov20}
	
	\begin{figure*}
		\centering
		\subfigure[4 hours]{
			\includegraphics[width=\columnwidth]{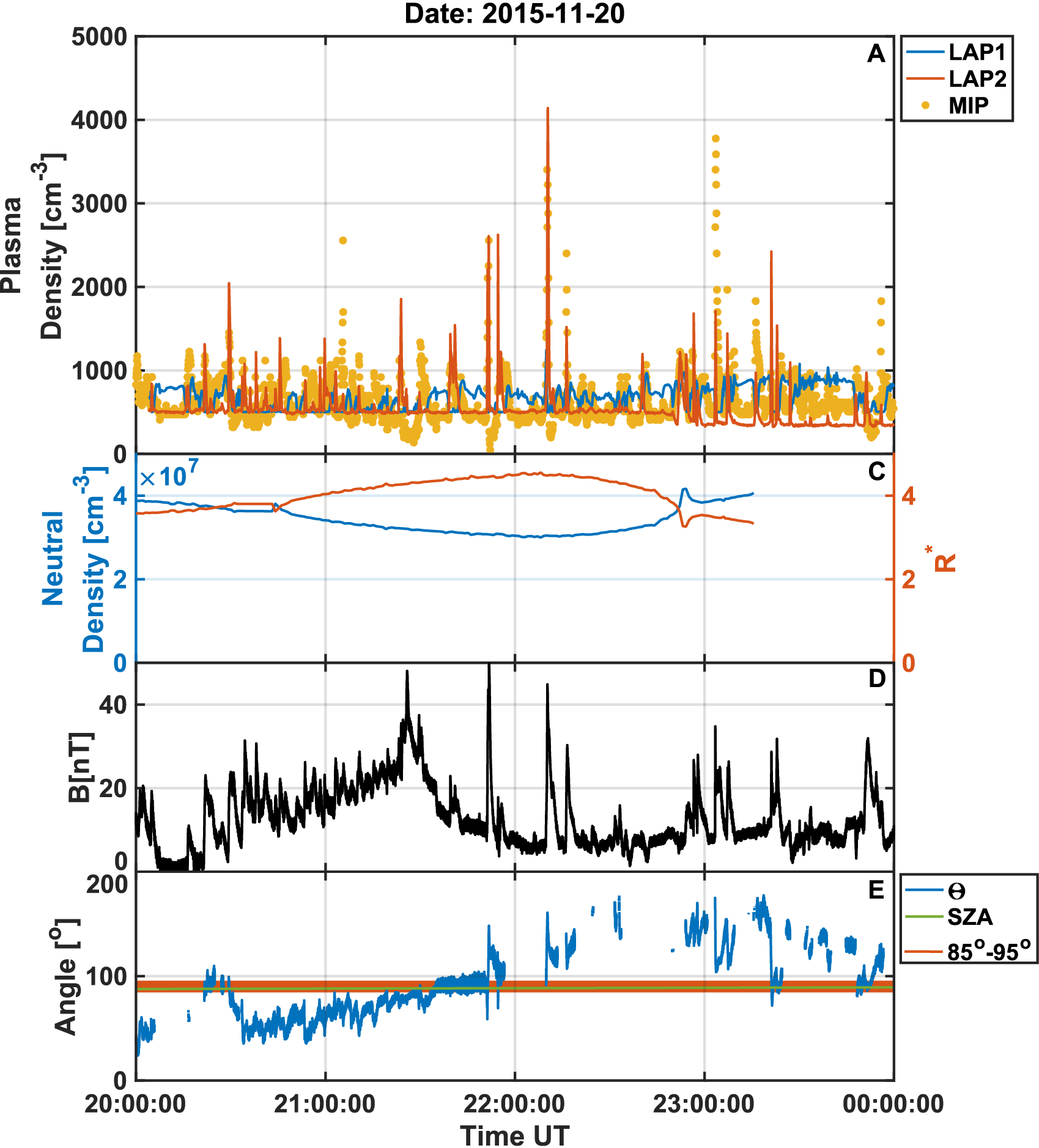}
			\label{fig:Nov201a}
		}
		\subfigure[1 hour]{
			\includegraphics[width=\columnwidth]{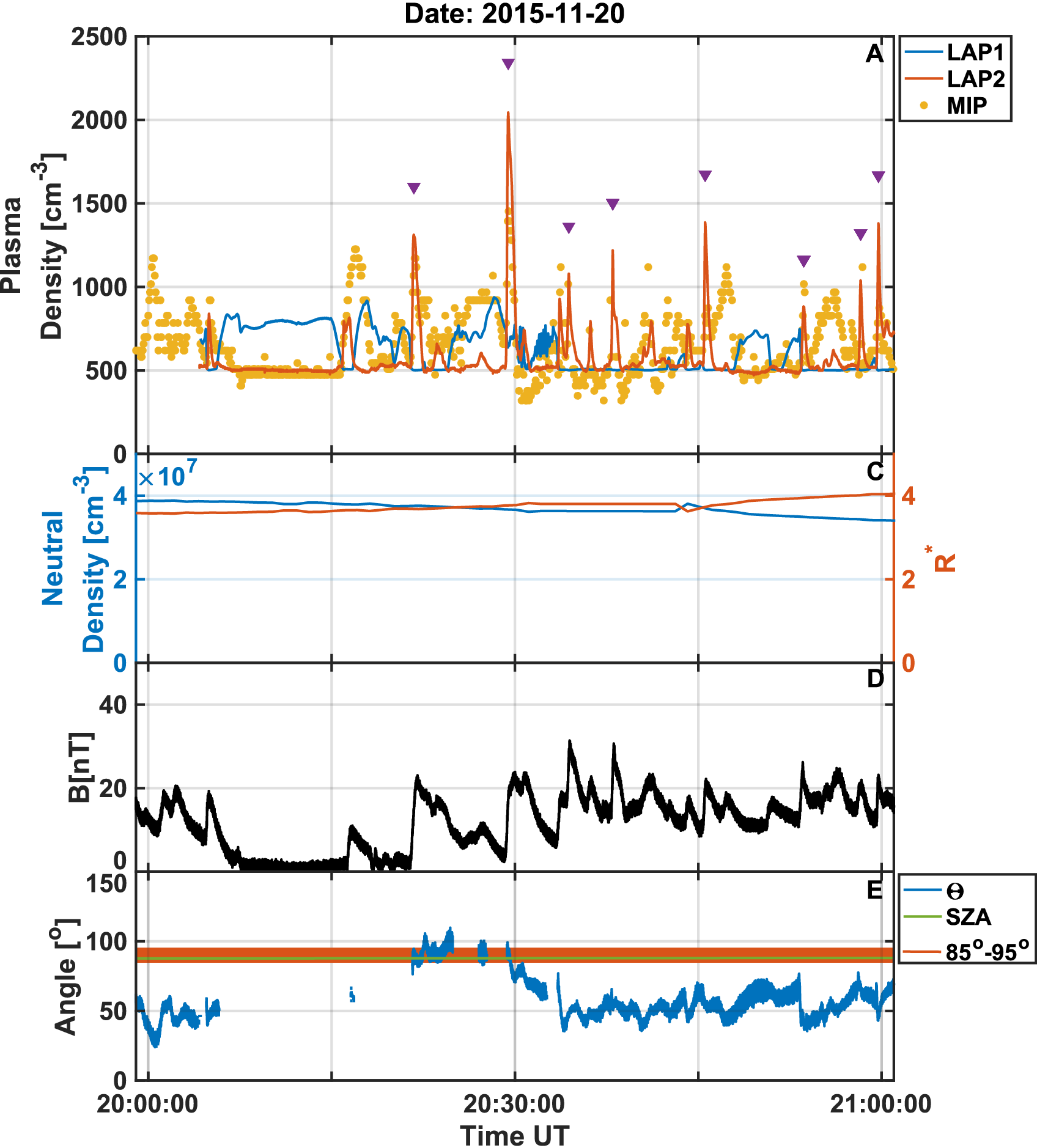}
			\label{fig:Nov201aH}
		}
		\caption{Fig. \ref{fig:Nov201a}: Overview data from November 20. LAP is in EI mode for this short macro block. Fig. \ref{fig:Nov201aH}: One hour of data from November 20, 22:00-23:00. The purple triangles show the pulses that have been detected with the automated algorithm used in section \ref{sec:boundary}. The panels for both figures are as described in Section \ref{sec:data}.}
	\end{figure*}
	
	Figure \ref{fig:Nov201a} shows the data from November 20. Rosetta was in this case close to the terminator on the morning side of the nucleus at about 150~km distance (Figure~\ref{fig:Orbit}). As can be seen in Panel~C, the neutral density is much higher in this event. This is not only because Rosetta is closer to the nucleus, but also in the southern hemisphere which during this time of the mission was in summer and thus more active \citep{Hansen2016}. Hence Rosetta is only at about 4 times the distance of the electron exobase. The plasma density is also much higher, resulting in good MIP density estimates during all of the day. As expected \citep{Henri2017} when close to the exobase, Rosetta sometimes is inside the diamagnetic cavity, with 82 cavity observations listed for this day by \citet{Goetz2016b}. 
	
	As in the event before (see Section~\ref{sec:oct24}), LAP is in the EI mode with LAP 1 at positive bias voltage to sample electrons and LAP 2 negative to sample ions. Interestingly, the densities measured by LAP 1 and LAP 2 seem to be anti-correlated. This is because the plasma density is higher on this day than October 24, which mainly is due to that Rosetta is closer to the nucleus and in the southern hemisphere. 
	At higher plasma density, the spacecraft potential is usually more negative, about -12~V \citep{Odelstad2015}. 
	If the spacecraft potential is more negative than the positive bias voltage applied to the probe, then the probe is not positive with respect to the plasma and electrons can not be collected. Therefore when the plasma density increases, the electron current to the probe can decrease, and therefore the density derived from LAP 1. The effect is stronger if the electrons inside the pulse are colder than outside \citep{Eriksson2017}, as they then will have even less kinetic energy to overcome the more negative $V_{sc}$. An example of direct measurement of more negative spacecraft potential in pulses will be shown in Section~\ref{sec:nov15}. 
	
	For a negatively biased probe, like LAP 2 in this case, a more negative spacecraft potential inside the pulses serves to increase the ion current. This adds to the direct increase of the ion current due to the higher density. The ion current measurements, at negative bias potential, therefore always correlate with the plasma density. We can see that in Panel~A, where MIP and LAP~2 density variations agree very well. Negatively biased probes are therefore safer to use for obtaining plasma density in a dense plasma. However, when the ion current is much lower than the electron current that is, when the plasma density is low, the signal to noise ratio in the ion current is low. Fortunately the spacecraft potential depends on density and will be low in such cases so the density variations can be observed in the electron current to a LAP probe at positive bias (like in the October 24 event, Section~\ref{sec:oct24}).
	
	Taking a look at a one hour interval, shown in figure \ref{fig:Nov201aH}, one sees that coincident pulses are seen in the magnetic field and plasma density. The pulses thus appear to be compressional in nature, as discussed by \citet{Hajra2017}. During this hour Rosetta was inside the diamagnetic cavity between 20:08 and 20:16 \citep{Goetz2016}. 
	The angle between B-field and nucleus direction is within about 45$^\circ$ of the perpendicular, and there is little change of this angle in the pulses.

	\subsection{November 15, 2015}
	\label{sec:nov15}
	
	\subsubsection{Morning}
	\label{sec:nov15am}
	
	\begin{figure*}
		\centering
		\subfigure[12 hour]{
			\includegraphics[width=\columnwidth]{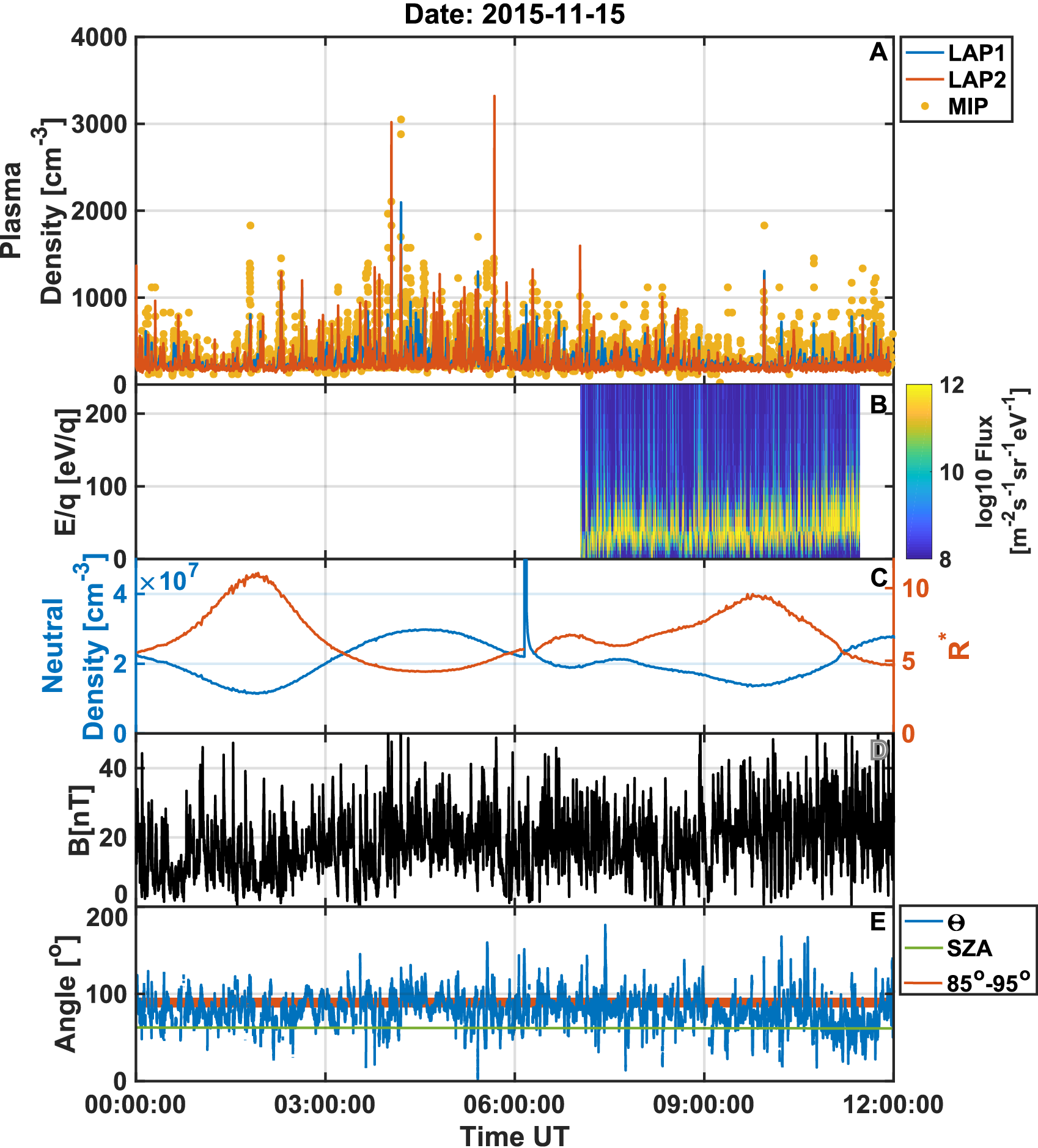}
			\label{fig:Nov151}
		}
		\subfigure[1 hour]{
			\includegraphics[width=\columnwidth]{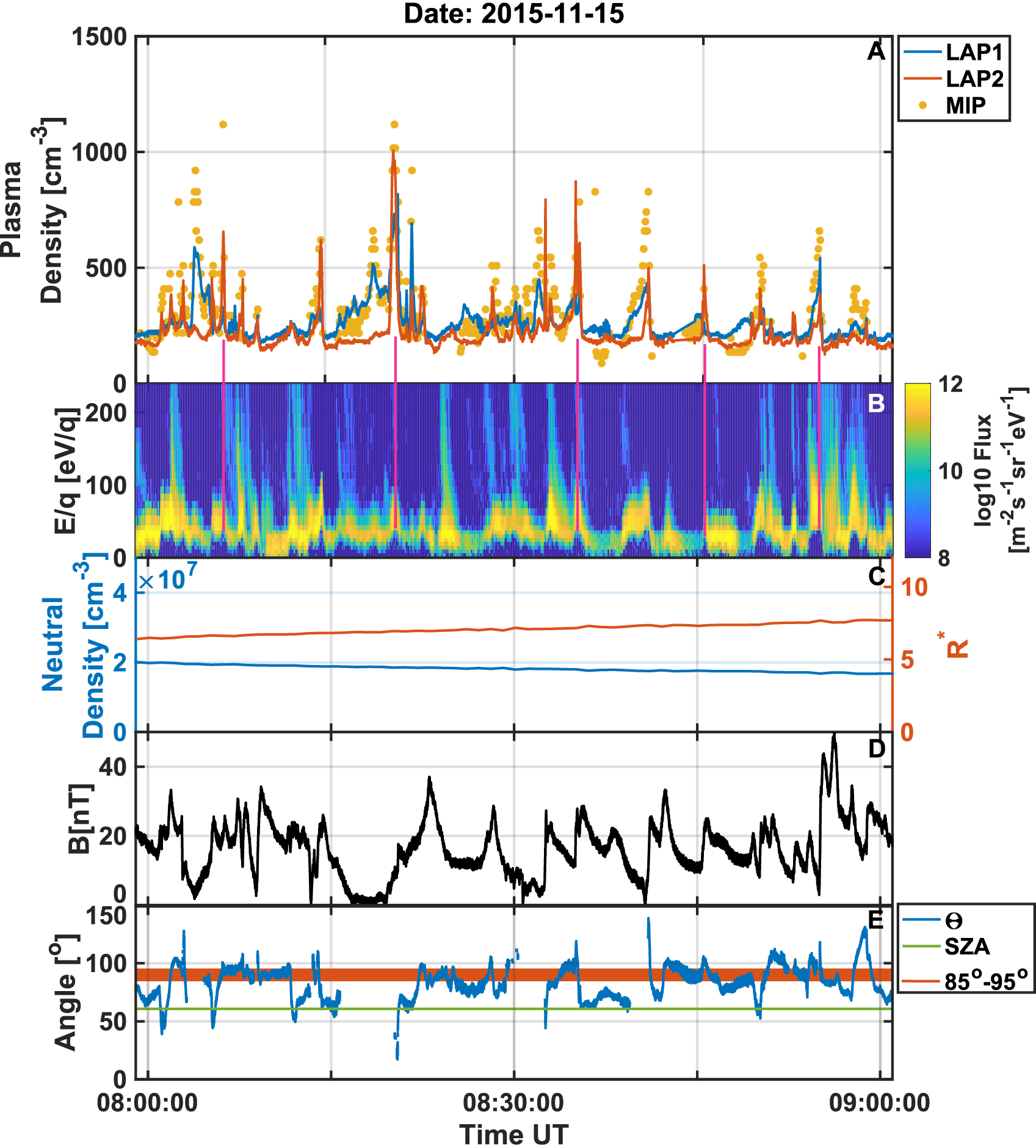}
			\label{fig:Nov151H}
		}
		\caption{Fig. \ref{fig:Nov151}: Overview data of November 15 in the morning. Here LAP is in II-mode. Fig. \ref{fig:Nov151H}: November 15 one hour (08:00-09:00). The panels for both figures are as described in Section \ref{sec:data}. Added are pink lines to show the correspondence between ICA's lower energy cutoff and the pulses.}
		\label{fig:Nov151_full}
	\end{figure*}
	
	November 15 is a special day as we have the first half of the day in II mode and the second half in VV mode, both with (some) coverage of ICA high resolution data. COPS neutral gas density data are basically identical for the two 12 hour intervals, and so the magnetic field seen by MAG and the plasma density by MIP are also quite similar, so the two macro blocks are comparable to each other. The VV data can be used to investigate how the spacecraft potential, $V_{sc}$ varies inside a pulse. 
	
	During this day Rosetta was close to the equatorial plane in CSEQ coordinates. The solar zenith angle is around 60$^\circ$ and the cometocentric distance about 150~km (Figure~\ref{fig:Orbit}). Figures \ref{fig:Nov151} and \ref{fig:Nov151H} show the full 12~hour interval and a zoom in to one hour where LAP operated macro II, measuring the ion density on both probes. Figure \ref{fig:Nov151} shows the familiar variations and pulses in the LAP probe density. The plasma density reported by MIP is much higher than for October 24th and about half the value of the November 20 event. Rosetta is about 5 times as far from the nucleus as the electron exobase and there are only two diamagnetic cavity sightings on this day, each of around 2~minutes around 05:50 and 08:18 \citep{Goetz2016b}.
	
	Figure \ref{fig:Nov151H} indicates that, despite operating identically, the LAP probes collect slightly different densities. It seems unlikely that this can be explained by very small scales of the plasma structures, as they last for several minutes and are expected to travel at least as fast as the neutral gas, or around 1~km/s \citep{Vigren2017}. More probably the difference is due to spacecraft attitude with LAP 2 being sometimes in the wake of the spacecraft, depending on the plasma flow direction, as discussed below. 
	
\begin{figure}
		\centering
		\includegraphics[width=\columnwidth]{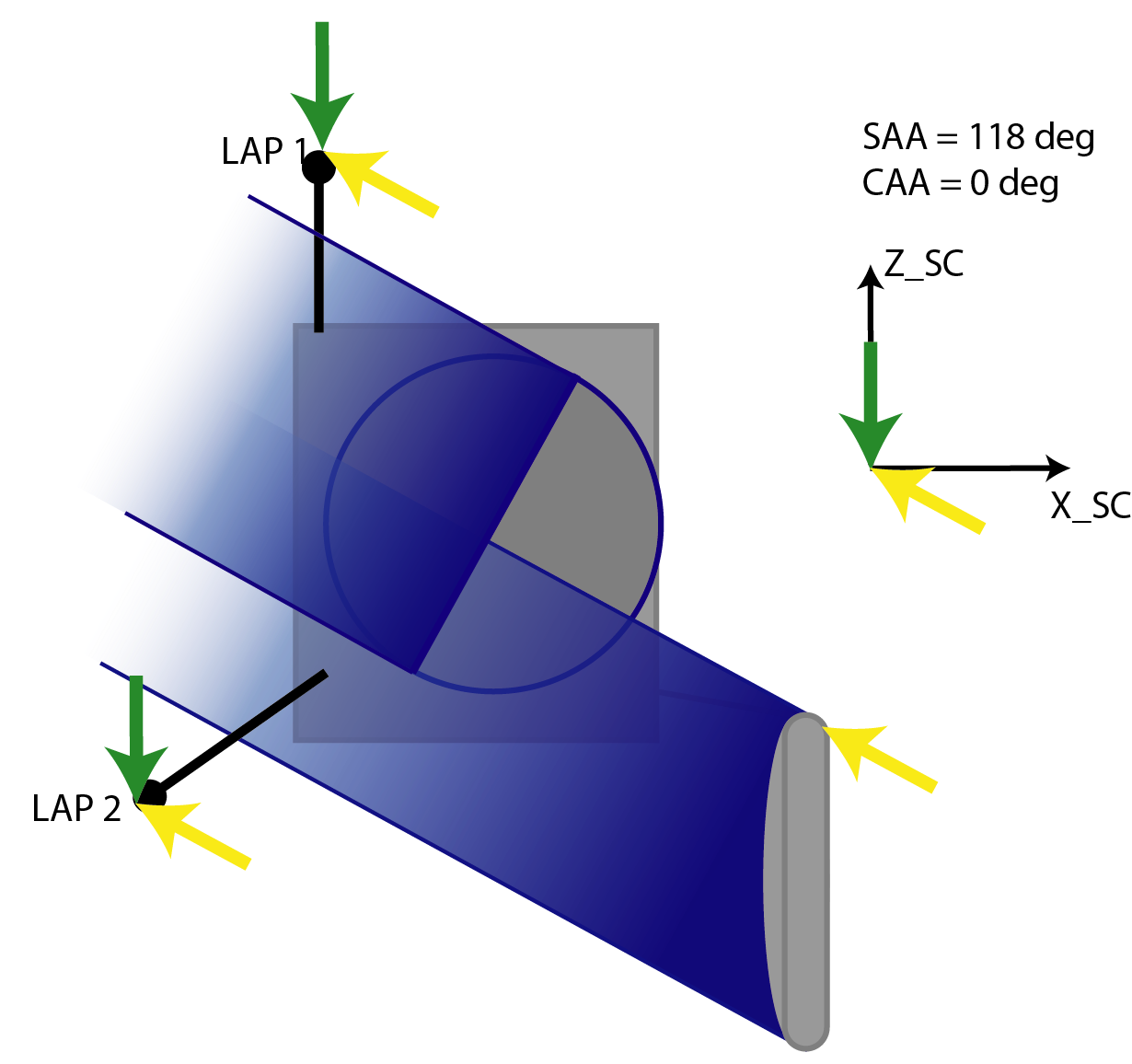}
		\caption{Rosetta pointing on November 15, 2015. The grey box is the spacecraft body and the grey object to the lower right is the high gain antenna. The solar panels extend into and out of the page (along the $Y$ axis) as well as LAP 1. The solar (SAA) and comet (CAA) aspect angles are the angles of the Sun and the nucleus with respect to the $+Z_{SC}$ axis. Yellow and green arrows denote the flow directions from the Sun and the comet nucleus, respectively. The shadows of the solar panels and the high gain antenna are indicated. 
Further explanations in the text.
		}
	\label{fig:pointing}
	\end{figure}

	Figure~\ref{fig:pointing} shows the pointing of Rosetta with respect to the comet nucleus and the Sun during this day. The yellow and green arrows illustrate the flow directions of solar photons and cometary ions, respectively, assuming the ions flow radially from the nucleus. Both these directions varied by less than $2^\circ$ during the day. The solar panels are always kept perpendicular to the solar direction, meaning their edges move along the blue circle. LAP 1 is mounted on a boom protruding diagonally out of the paper \citep{Eriksson2006}, and can be in shadow only if it is behind the solar panels, which it never is in this event. The line from the Sun to LAP 2 could on the other hand be blocked either by the spacecraft body or by the high gain antenna (seen at lower right), but we can see this does not happen here. Therefore both probes are sunlit during all this day.

	There could also be a wake effect because of a radial ion flow from the comet (green arrows). It can be seen that here is no obstacle directly blocking the flow to any of the probes in this case. However, while LAP 1 is on a boom protruding toward the nucleus and should experience an unperturbed plasma flow as long as the radial component is directed away from the nucleus, LAP 2 could experience some kind of bow wave or similar extending from the forward edge of the spacecraft. Studying Figure~\ref{fig:pointing}, one can see that a turn of the flow (green arrow) by about $20^\circ$ in the clockwise direction would be sufficient to bring LAP 2 into a sharply defined wake with edges parallel to the flow direction. In reality, a region roughly defined by a Mach cone from the spacecraft edges will be perturbed \citep{Hastings1995}, so LAP 2 may very well see such disturbances. That the plasma flow may depart from the exactly radial direction from the nucleus further increases the possibility for wake effects on LAP 2. A few nucleus radii away from the nucleus, kinetic and hydrodynamic models agree that the neutral gas flow must be close to radial from the nucleus \citep{Tenishev2008,Bieler2015} but the coupling of gas and plasma is expected to be far from perfect \citep{Vigren2017}, particularly at small scales. That this coupling is not perfect at least at small scale is obvious if comparing the large plasma density variations seen in Panels~A to the more smooth neutral density in panels~C. Wake effects and changing flow directions are possible sources for the differences between the two LAP probes in panel~A of Figure~\ref{fig:Nov151H}, particularly as LAP 2 usually observes the lower density. 
	
	For this event ICA high resolution data are available (Figure~\ref{fig:Nov151H}). The density pulses detected by LAP can be seen to correspond to ICA increases in ion flux and ion energy (B). The ion energy increase in a pulse can at least partly be explained by the more negative spacecraft potential, accelerating ions toward the spacecraft. The ion flux should increase with the density. When looking in more detail, it appears that there is a good correspondence between increases in ICA's lower energy cutoff (lower edge of yellow region, see pink lines for examples) and LAP density pulses, as it is expected for ions accelerated by the spacecraft potential that become more negative with increasing density (a few examples are indicated by vertical lines).
	However, the ICA ion flux can intensify or spread to higher energy with little or no corresponding density increase (LAP density pulse). Examples of this can be seen around 08:24 and 08:47. Such signatures can be interpreted as ions accelerated at some distance from the spacecraft and now reaching it with no or moderate density increase. These share the characteristics of type 5 in the classification of short lived ICA ion features by \citet{StenbergWieser2017}. 
	
	As in the previous events, the magnetic field shows strong signatures coincident with the plasma density and ion flux enhancements. Also as in previous events, the magnetic field direction between pulses is at a quite large angle to the direction to the nucleus, but the pattern at the pulses it more alignes with the nucleus direction. This is similar to the type 5 signatures in ICA \citep{StenbergWieser2017}

	\subsubsection{Afternoon}
	\label{sec:nov15pm}
	
	\begin{figure*}
		\centering
		\subfigure[12 hour]{
			\includegraphics[width=\columnwidth]{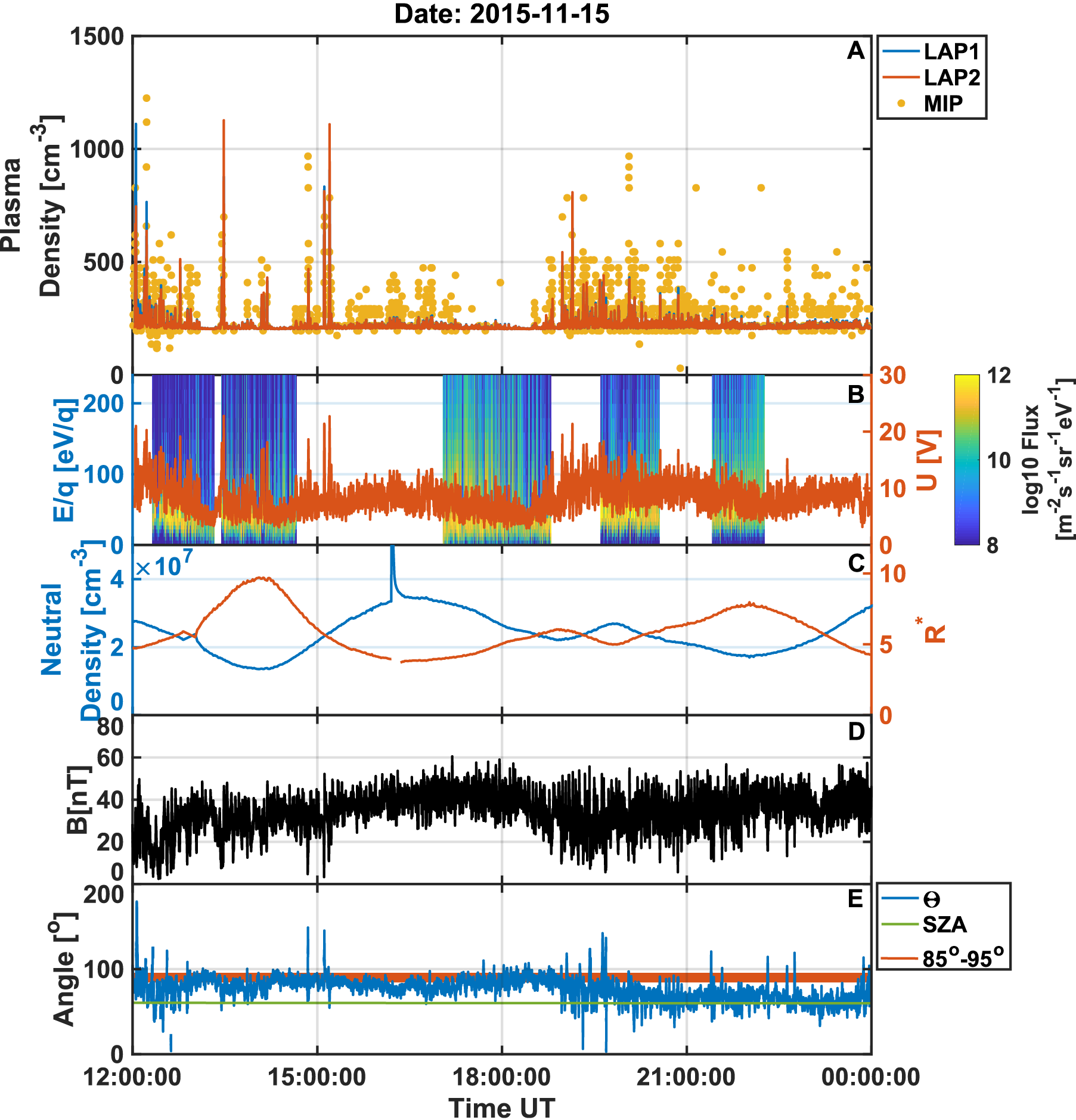}
			\label{fig:Nov152}
		}
		\subfigure[1 hour]{
			\includegraphics[width=\columnwidth]{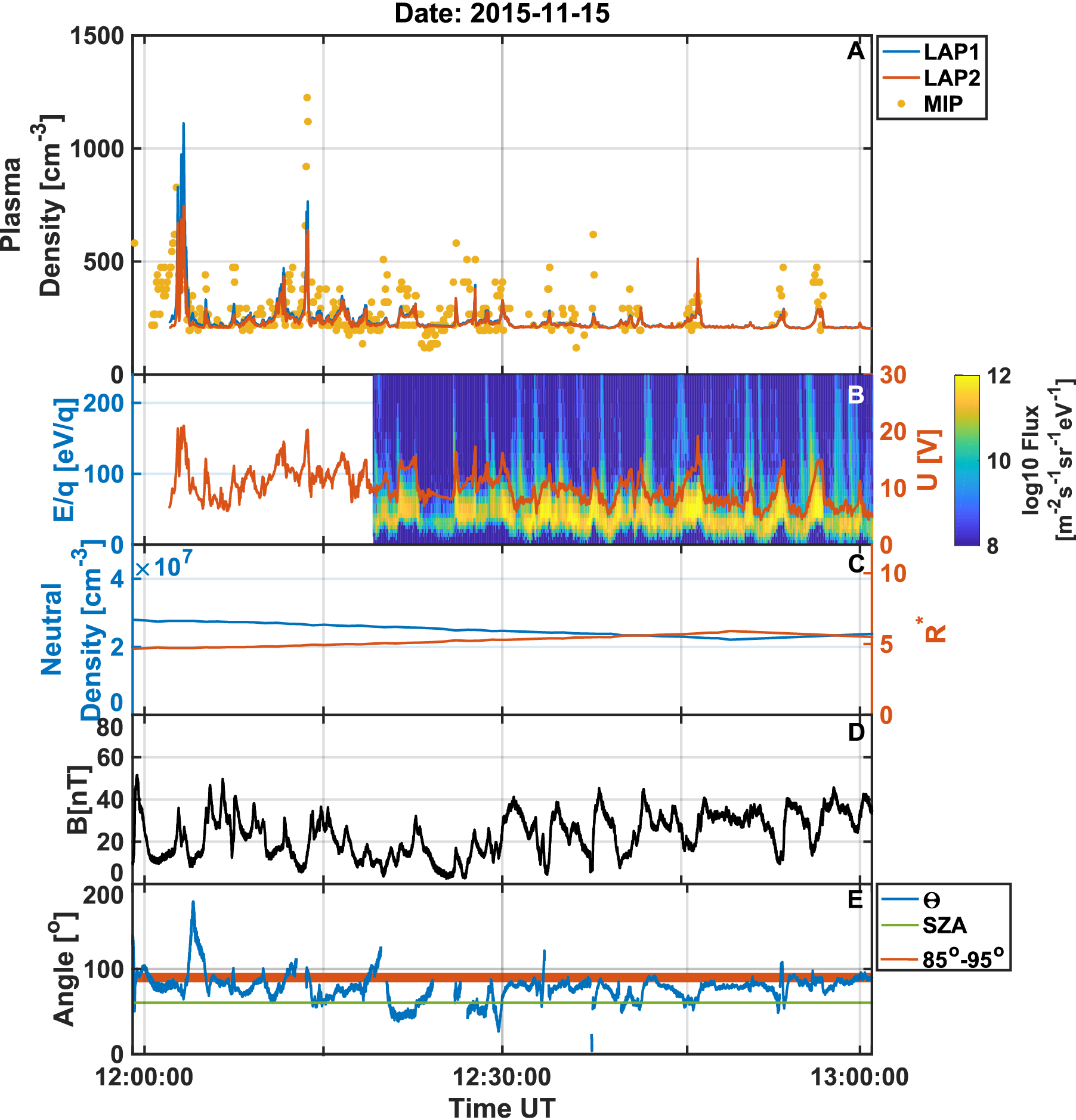}
			\label{fig:Nov152H}
		}
		\caption{Fig. \ref{fig:Nov152} Second part of November 15. MIP, LAP in VV mode. ICA, COPS and MAG show similar plasma dynamics to the previous part of the day. Thus these are comparable to \ref{fig:Nov151}. Since LAP is in E-field mode producing voltages, it is shown in panel~B together with the ICA measurements with energy scale in eV, with some stretching of the voltage range for clarity. There are data gaps in ICA until 12:18, 14:50-17:00, 18:50-19:40, 20:15-21:20 and 22:20-23:10 
		Fig. \ref{fig:Nov152H}: One hour of November 15 (12:00-13:00). The panels for both figures are as described in Section \ref{sec:data}.}
		\label{fig:Nov152_full}
	\end{figure*}
	
	During the second half of this day, LAP operated both probes in floating mode (VV), as shown in Figures~\ref{fig:Nov152} and \ref{fig:Nov152H}. The quantity measured here is the probe potential with respect to the spacecraft body, which gives an estimate of the negative of the spacecraft potential \citep{Odelstad2017}. 
	Comparing the data from MIP (Panel~A), ICA (B), COPS (C) and MAG (D) to what we found in the first half of the day (Figures~\ref{fig:Nov151} and \ref{fig:Nov151H}) we find similar signatures. In particular, Panel~B shows that the pulses in ICA spectra are seen in LAP both in probe currents (in the morning) and in the probe voltage (afternoon). The LAP 1 and LAP 2 voltages calibrated to MIP plasma density data as described in Section~\ref{sec:calibration} are shown together with the MIP data in Panel~A. We clearly see the spacecraft going more negative in the pulses of higher density as suggested in Sections~\ref{sec:nov20} and \ref{sec:nov15am}. The two LAP probes observe very similar potentials when in this mode, while the probe currents measured in the II mode were seen to differ more between the probes. This can be understood as the voltage picked up by the probes in VV mode being dominated by the potential of the spacecraft as a whole, which is a common property of both probes, while the currents in II mode more strongly depends on the local plasma conditions at each probe, which as noted in Section~\ref{sec:nov20} can be influenced by for example wake effects. Nevertheless, there is some difference between the two probe voltages. This difference was used by \citet{Karlsson2017} to derive the electric wave field in this particular event, finding waves in the lower hybrid frequency range on the edges of the pulses. 	
	
	\subsection{Distribution with Distance}
	\label{sec:boundary}
	
	\begin{figure*}
		\centering
		\subfigure[]{
			\includegraphics[width=\columnwidth]{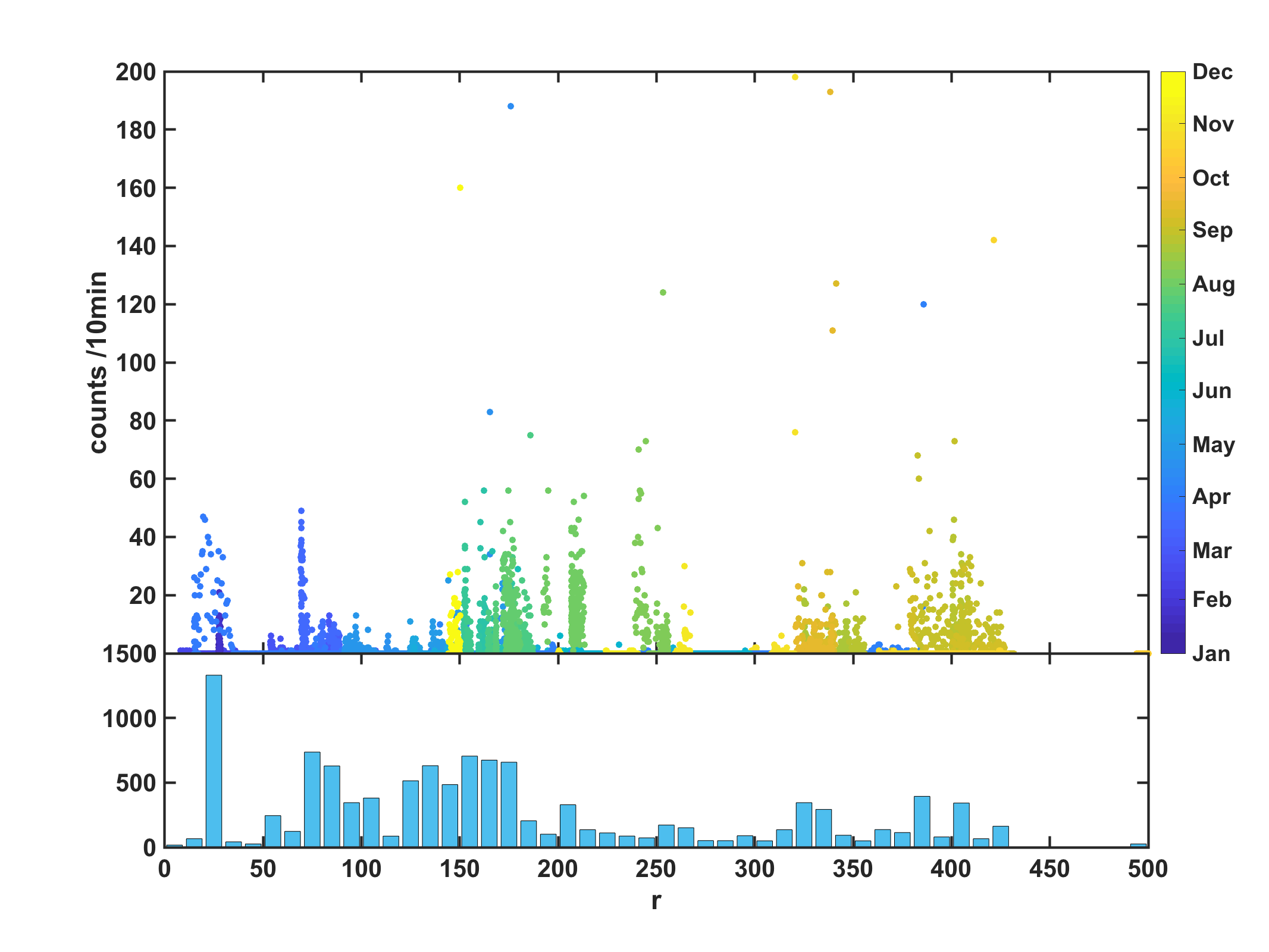}
			\label{fig:rRadz}
		}
		\subfigure[]{
			\includegraphics[width=\columnwidth]{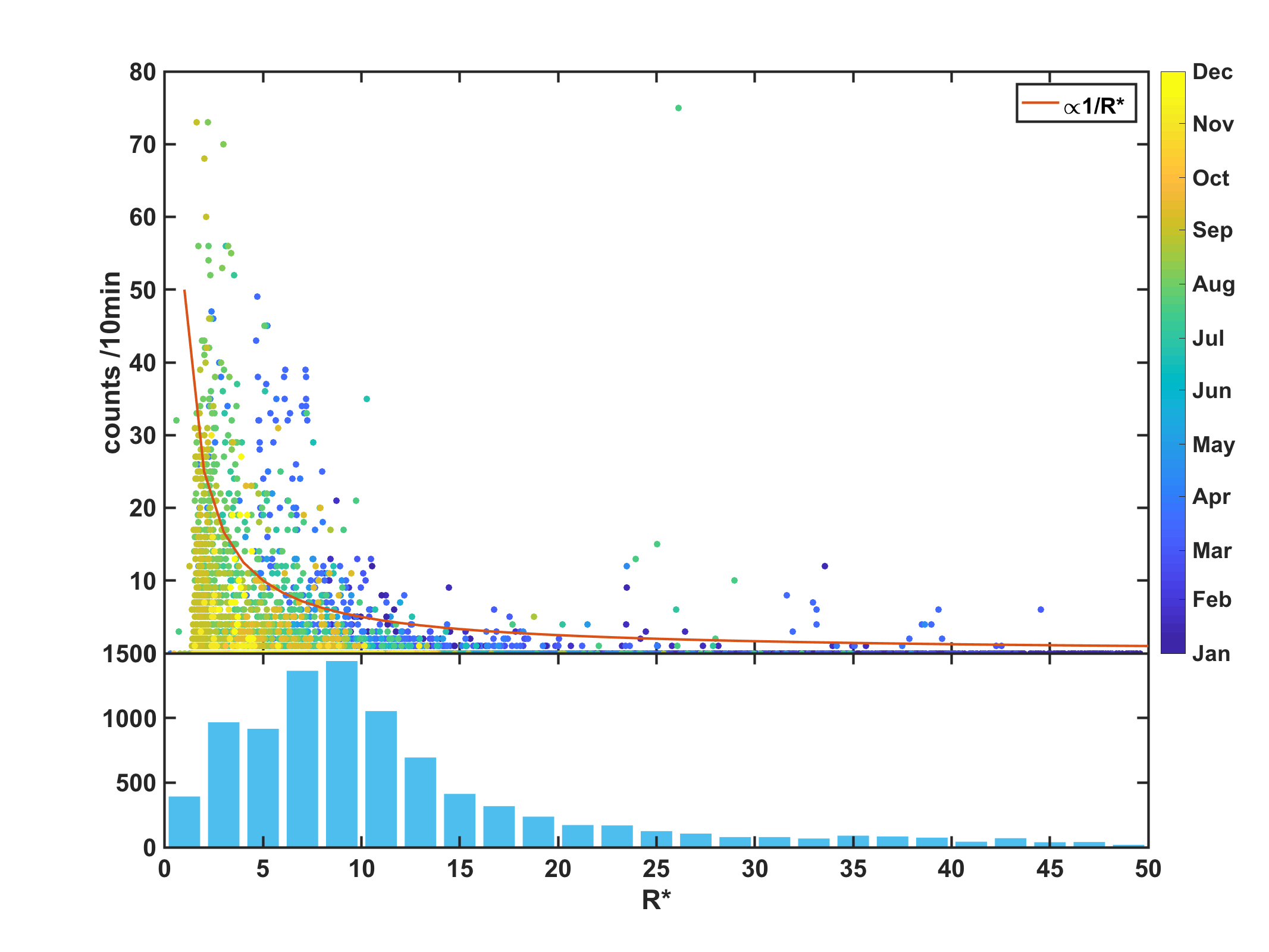}
			\label{fig:rstarz}
		}
	\subfigure[]{
		\includegraphics[width=\columnwidth]{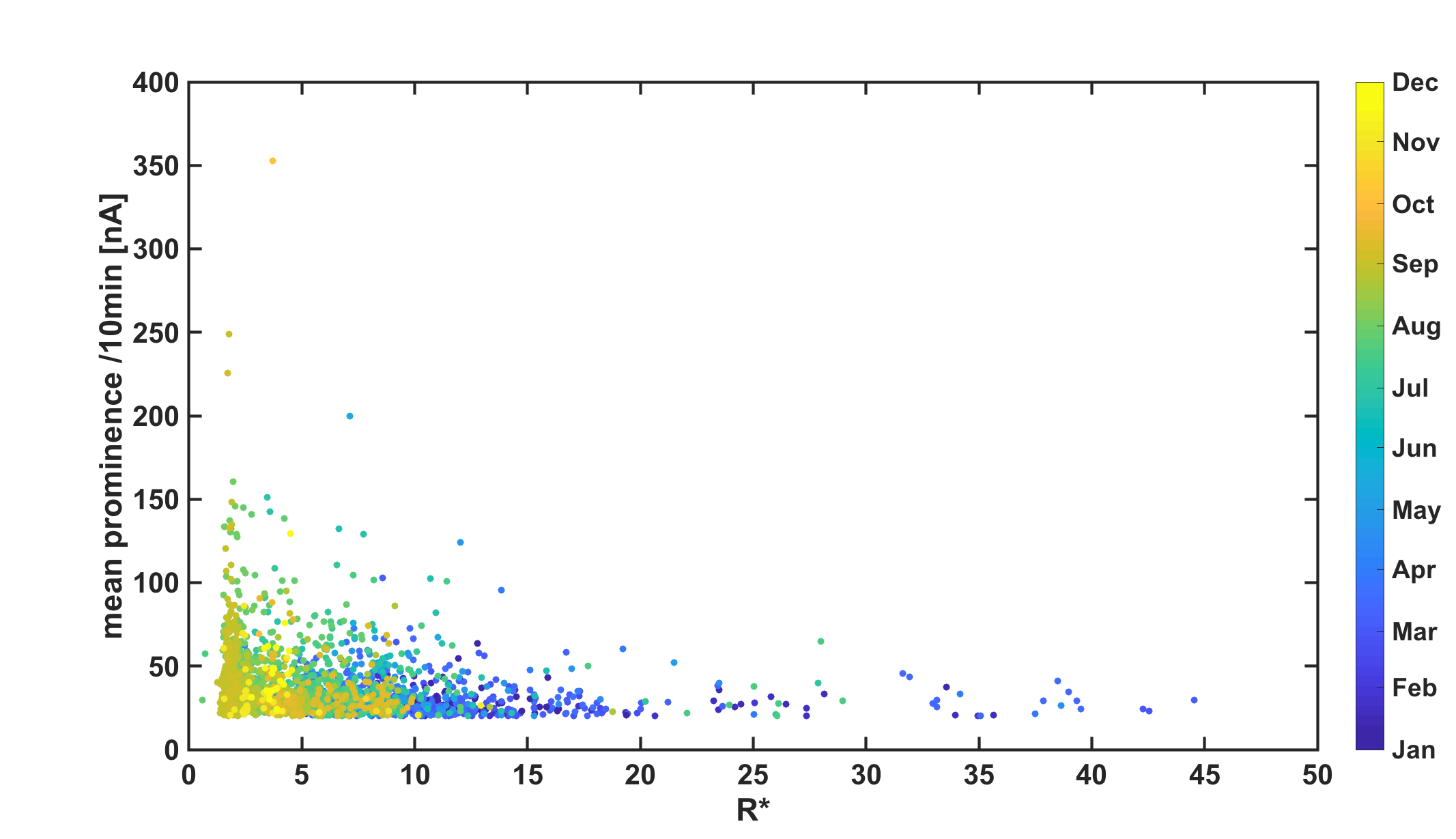}
		\label{fig:rstarzprom}
	}
	\subfigure[]{
		\includegraphics[width=\columnwidth]{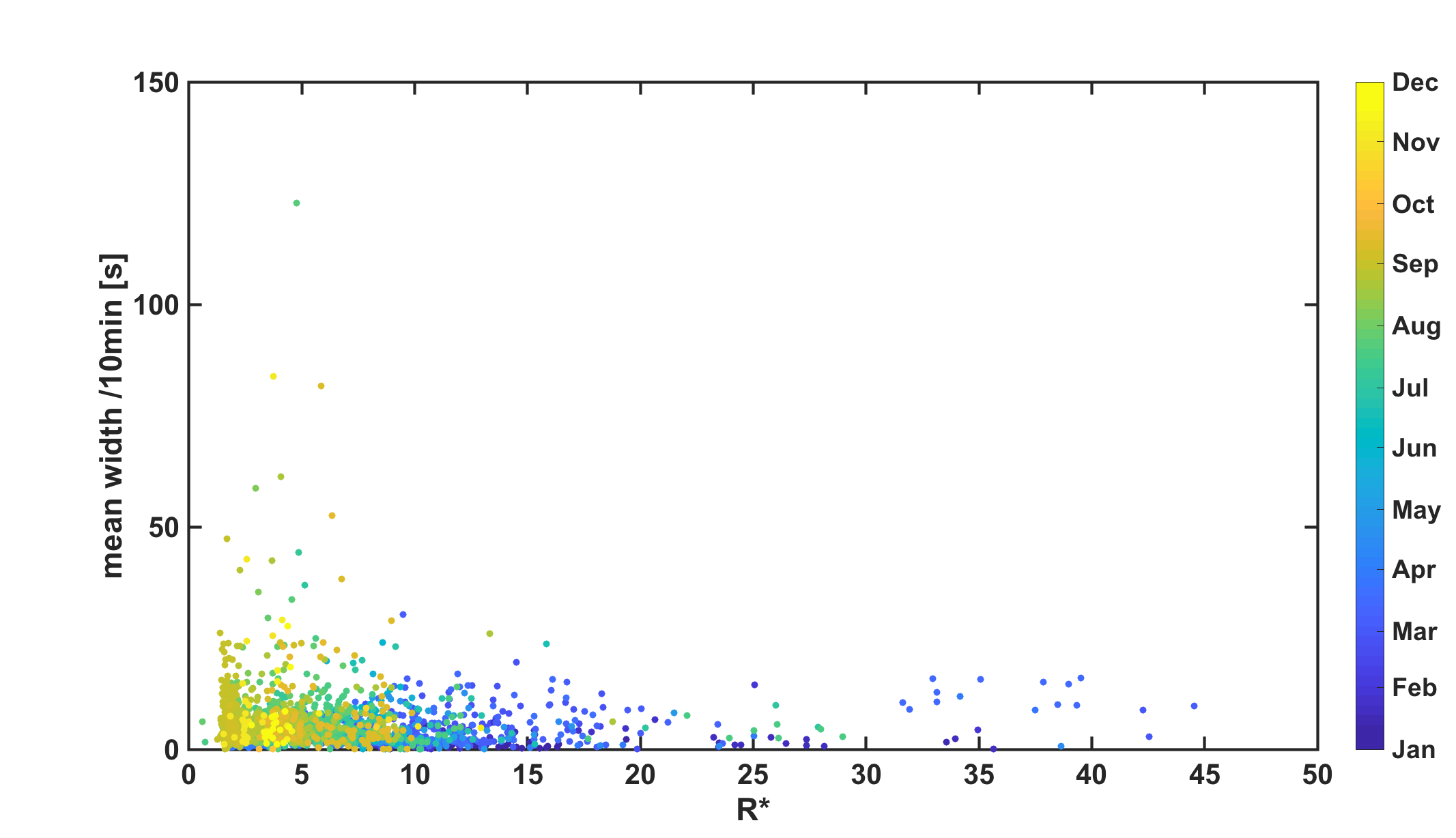}
		\label{fig:rstarzwidth}
	}
		\caption{Fig. \ref{fig:rRadz} shows the number of detected pulses, counts, during a 10 minute interval of data from LAP2. This is plotted against the radial distance of Rosetta. No clear correlation visible. The lower panel shows the amount of 10 minute intervals spent at a specific radial distance interval. Fig. \ref{fig:rstarz} is again the number of counts during a 10 minute interval but plotted against the distance of Rosetta with respect to the electron cooling boundary, $R^*$. We see that the data is sorted and decays approximately as 1/R. The red line gives an a 1/R dependence. Most of the data lies where Rosetta is at between 6 and 10 times the cometocentric distance of the exobase. 
		Figure \ref{fig:rstarzprom} and \ref{fig:rstarzwidth} show the prominence and width of the pulses as sorted by the distance to the exobase.}
	\end{figure*}
	
	It can be seen from the events above that the pulses can be found at various distances from the nucleus, both absolute distance in km and distance relative to the electron cooling boundary. \citet{Henri2017} showed that the relative distance to this boundary well organized observations of the diamagnetic cavity. As the definition of the cooling boundary distance depends on the cometary activity $Q$ we may expect this to bring some order also to the pulse observations.
	
	To investigate this statistically, we have used ion current data from LAP 2 for the full year 2015. To get comparable data, we only use data from two macros with good distribution over the year, known as 525 and 624, both with LAP 2 measuring ion current. The burst mode data (624) has a sampling frequency of 58.7~Hz while the normal mode data (525) is recorded with a frequency of 0.45~Hz. As the typical pulse duration is several tens of seconds also the lower sampling rate is sufficient. For this purpose we have not calibrated the data to MIP densities but used the raw collected LAP current, which is proportional to the plasma density. The data set is similar to what was presented in \citet[Fig.~6]{Eriksson2017} but the pulse finding algorithm is slightly refined. The present algorithm uses a threshold for the prominence (the amplitude over a background) of a peak in the LAP 2 ion current. Given any signal peak (local maximum), the prominence is defined as the difference of this peak value and the minimum value found between the peak and the nearest point where the signal reaches at least the same magnitude (or, if it never does, an end point of the 10 minute data interval). One such prominence value can be defined before the peak and one after the peak, and the smallest of the two is used. The plot of occurrence statistics as function of mission date and cometary longitude provided by \citet[Fig.~6]{Eriksson2017} does not change appreciably by this, but the new method avoids some problems when pulses occur close together. An example of how the pulse detection works is shown in Figure~\ref{fig:Nov201aH}, where pulses found by this criterion are marked by purple triangles. The threshold set for the peak prominence is 20~nA, which for this event corresponds to a density increase of about 400~cm$^{-3}$ (Table~\ref{tab:LapFits}).
	
	We take the number of pulses detected in a non overlapping 10 minute interval and plot versus where it was observed. Figures \ref{fig:rRadz} and \ref{fig:rstarz} show pulse counts versus the radial distance r and versus the position of Rosetta with respect to the electron cooling boundary, $R^*$, respectively. The histogram below each plot shows the amount of data points in a position bin.
	The amount of pulses in a 10 minute interval is not well sorted by the radial distance, (Fig. \ref{fig:rRadz}). However, from Figure~\ref{fig:rstarz} we see that the data is much better organized by the position relative to the electron exobase. For comparison, a $1/R^*$ curve is also shown (red line). Rosetta spends most of its time at about 5-10 times the cometocentric distance of the exobase, peaking at 9, but the higher number of pulses are mostly seen closer to the nucleus. 
	
	In Figure~\ref{fig:rstarzprom} is shown the mean prominence of the peaks found in the same 10~minute intervals as above. Only plots sorted by $R^*$ are shown, as the radial distance $r$ in km did not sort the data well in this case either. As discussed above, the prominence is the amplitude of the pulse compared to the background value around it. This is LAP probe current rather than calibrated density, but as the current is proportional to the density we can see that the highest pulses occur closer to the exobase. Figure~\ref{fig:rstarzwidth} show the same characteristics for the width in seconds, which is also well sorted by $R^*$ but not by $r$ (the latter not shown). The pulse width varies, but typically lies between a few seconds and a few tens of seconds. Broader pulses were mainly detected when the comet was most active, during August-December, which also is the period of the events presented above.
	
	\section{Discussion and Conclusion}

	\subsection{Summary of observations}
	
	We have shown RPC data for four events, during the months after perihelion. The events show varying radial distance to the nucleus, phase angle and local time. These events all show pulse-like intensifications of plasma density, ion energy and flux, and magnetic field intensity. The enhancement of density and magnetic field is often very strong, up to an order of magnitude and sometimes even more. The plasma density increase in a pulse is seen in LAP ion and electron currents as well as in MIP plasma frequency data, ICA ion flux and LAP spacecraft potential. The magnetometer timeseries is smoother than the other data, which can be understood from Biot-Savart's law. The plasma measurements are local while the magnetic field is an integration of the current density over a large volume.
	
	We could also observe the spacecraft potential going more negative within a pulse, but combining density from MIP plasma frequency with ion and electron currents from LAP shows that the LAP current intensifications are no artifacts of varying spacecraft potential. Instead the varying spacecraft potential can be interpreted as varying plasma density. However, in dense plasmas, when the spacecraft potential is sufficiently negative, the response of the LAP electron current to a density intensification is complicated by the spacecraft potential change and LAP electron current measurements are not always reliable in this situation (Section~\ref{sec:nov20}). Using the LAP ion current, calibrated to MIP density from plasma frequency determination, removes this ambiguity. Pulses are generally seen in both probes (both measuring ion current or one measures electron current and the other ion current) but they do not look exactly the same even when the probes are identically operated (see for example figure \ref{fig:Nov151_full}). This may be due to different plasma conditions at the two probes because of effects of the spacecraft on the plasma, like the formation of a wake and even a bow wave.
	
Using the neutral gas number density from COPS we could define the cometocentric distance of the electron exobase and normalize the cometocentric distance to this. In the events studied here Rosetta was at 3-10 the times distance of the electron cooling boundary. The full year statistics showed pulse observations over a larger range of distances, but few are found outside 20 times the height of the exobase. In general, more pulses are found close to the exobase than far away. 

The distribution of pulse observations in Figure~\ref{fig:rstarz} is in shape similar to what \citet{Henri2017} found for diamagnetic cavity observations, but the pulses are seen further out than the cavity. The plasma density in the cavity is much smoother than the region outside, though density pulses have been found also in about 15\% of all cavity events \citep{Hajra2017}. This is consistent with that we only find a few pulses that are situated inside 2 $R^*$. In absolute values of the distance normalized to the exobase distance, $R^*$, the cavity observations are confined within about $R^* < 5$, while the pulses we observe are frequently seen at least to $R^* = 20$ in our statistics. Furthermore, we find the pulses of highest amplitude and width closest to the electron exobase. Our statistics are based on a threshold for how much the magnitude of a current pulse should rise over the background. That we find pulses to have lower amplitude far away will skew the occurrence statistics in the sense that there may be many but smaller pulses at large distance, where some might be missed due to the set threshold. While the electron exobase scaling fits well, there could still be some other process scaling in a similar way, for example ion collisionality, which regulates the stability of the cometary ionosphere.

\subsection{Comparison to simulation results}
	
	The observed plasma variations can be compared to the global 3D hybrid simulation model by \citet{Koenders2015} who studied the cometary plasma at an activity of $Q = 5 \cdot 10^{27}$~s$^{-1}$. Table~\ref{tab:cases} shows that this is the relevant range for the events we present. In the hybrid simulation \citep[Figures~3 and 6]{Koenders2015}, filaments or blobs of high density plasma were seen to detach from the diamagnetic cavity and move outward and ultimately tail ward. The density in these simulated pulses seems to reach above the value just inside the cavity by a factor of 2-5, and sometimes more than an order of magnitude over the density seen adjacent to the pulses. These pulses were seen also in the magnetic field strength which about doubled in the simulated pulses, though the phasing of the density and magnetic field increases were not always perfect. This corresponds well to what we observed in the events presented in this paper, but there are also differences. The plasma density seen in the simulation just inside the cavity in the plane through the nucleus perpendicular to the interplanetary magnetic field was around 5,000~cm$^{-3}$ while none of our events show much more than 1,000~cm$^{-3}$ in the cavity. Also the diamagnetic cavity in the simulation extended at most about 50~km from the nucleus, while in our events as well as in all cavity observations by \citet{Goetz2016b} it is seen to be much bigger. From \citet{Koenders2016a} the typical duration of high density pulses also seems to be from a few to about 20 seconds. Similar results are seen in the MHD models by \citet{Rubin2012}. This is mainly comparable to our results, but we also have many examples of wider pulses. It should also be noted that nothing like the "fingers" of unmagnetized plasma stretching out from the diamagnetic cavity that were inferred by \citet{Henri2017} has been reported from the hybrid simulations. The simulated cavity boundary was not perfectly smooth, but the variations quite small \citet[Figure~6]{Koenders2015}.
	
	The hybrid simulations also suggest the diamagnetic cavity is most unstable in the plane containing the nucleus and the interplanetary magnetic field (IMF), which we can call the magnetic equatorial plane. We do not have access to the interplanetary magnetic field as the magnetic field at Rosetta is heavily influenced by mass loading and other cometary processes. Using the local magnetic field no clear confinement to the magnetic equatorial plane could be found in the events. However this could be due to the draping changing the direction of the magnetic field. The pulses could still be more prominent in the real magnetic equatorial plane than elsewhere as there is clear draping during the months around perihelion \citep{Goetz2016b}.
		
	Using the parameters of the simulation of \citet{Koenders2015}, the nominal exobase distance is 20~km. However the filamentation in the simulation starts at about 50~km which is at the cavity boundary. Our statistics suggest the filamentation start at the $R^*=1$ which according to \citet{Henri2017} is approximately the same as the cavity boundary at 67P. Our results are therefore consistent with the filamentation starting at the cavity boundary as seen in the simulations. Note that we use the same definition of the exobase for the simulation and the measured data. So the comparison is meaningful even though the exobase is not a well defined boundary.
		
	 From these points, it appears that the hybrid simulations capture many but not all of the features of our observed pulses, and also some but not the full physics of the diamagnetic cavity and its surroundings. A clue to the missing physics could be how the distance relative to the electron exobase is observed to organize observations both of the diamagnetic cavity \citep{Henri2017} and the pulses outside the cavity (this work). The exobase is the characteristic distance where electron collisions is no longer efficient, and therefore electron kinetic effects become important, which are missing in the hybrid simulations. To fully include such effects also kinetic electrons must be included, like in the particle in cell simulations presented by \citet{Deca2017}, which however lack collisional processes. 
	 Such simulations with spatial resolution of 10~km or better will be needed to resolve the diamagnetic cavity and its dynamics, and will when available presumably shed more light on the density and magnetic field pulses presented here.
	
	\subsection{Concluding remarks}
	We have shown that the localized density enhancements reported by \citet{Eriksson2017} are common around comet 67P. Furthermore they coincide with enhancement in magnetic field and ion flux. These characteristics and their distribution in space are at least qualitatively similar to filaments emanating from the diamagnetic cavity in the hybrid simulations by \citet{Koenders2015}. 
	
	This study also leaves some unanswered questions. Among these are: \\
	We see enhanced ion fluxes up to several hundred eV as reported by \citet{StenbergWieser2017} coinciding with the density and magnetic field pulses. The acceleration mechanism is unclear. 
	The electron temperature in the pulses needs to be investigated as \citet{Eriksson2017} shows only one example. We saw an indication in Section \ref{sec:nov15pm} that the temperature is different. 
	Another issue to resolve is if there are distinct types of pulses. The detailed examples were taken around perihelion suggesting only one kind of pulses. This can however be different at other times. \citet{StenbergWieser2017} found 5 different types of short lived ion flux enhancements. These types could reflect in types of density structures, for example with or without cold electrons reported by \citet{Eriksson2017}. \citet{Hajra2017} found that some pulses propagate inside the diamagnetic cavity. This could point to a class of pulses due to inward propagating waves reaching the cavity. Our statistics show that fewer pulses are found far out from the nucleus, suggesting most pulses originate from the inner region. Further studying the cold electrons in the pulses could show the source region of individual pulses since pulses containing cold electrons must come from the near nucleus environment, where cooling is efficient enough.

	
	\section*{Acknowledgments}
	
	Rosetta is a European Space Agency (ESA) mission with contributions from its member states and the National Aeronautics and Space Administration (NASA). This work has made use of the AMDA and RPC Quick look database, provided by a collaboration between the Centre de Donn\'{e}s de la Physique des Plasmas (CDPP) (supported by CNRS, CNES, Observatoire de Paris and Universit\'{e} Paul Sabatier, Toulouse), and Imperial College London (supported by the UK Science and Technology Facilities Council). The research in this paper was funded by the Swedish National Space Board under contracts 171/12 and 109/12. Work at LPC2E/CNRS was supported by ESEP, CNES and by ANR under the financial agreement ANR-15-CE31-0009-01.
	
	
	
	
	\bibliographystyle{mnras}
	\bibliography{ManExtracted}
	
	
	
	


	\bsp	
	\label{lastpage}
\end{document}